\documentclass[final,12pt]{amsart}
\usepackage[style=alphabetic,url=false, isbn=false, doi=false,maxbibnames=99]{biblatex}
\AtEveryBibitem{
    \clearfield{urlyear}
    \clearfield{urlmonth}
}
\usepackage{amsmath,amsfonts,amsthm,amssymb}
\usepackage{nicefrac,fancybox}
\usepackage{tikz}
\usepackage{wrapfig}
\tikzset{
	MyPersp/.style={scale=1.8,x={(-0.8cm,0cm)},y={(0cm,.3cm)},
    z={(0cm,1cm)}},
	MyPoints/.style={fill=white,draw=black,thick}
		}
\usepackage{tikz-cd}
\usepackage[nomargin,inline]{fixme}
\usepackage{tqft}
\usepackage{multirow}
\usepackage[margin=1in]{geometry}
\fxsetup{theme=color}
\author{Ben Webster}
\address[B.W.]{Department of Pure Mathematics, University of Waterloo \& Perimeter Institute for Theoretical Physics,
Waterloo, ON, Canada}
\email{ben.webster@uwaterloo.ca}
\author{Philsang Yoo}
\address[P.Y.]{Department of Mathematical Sciences \& Research Institute of Mathematics,
Seoul National University, Seoul, Republic of Korea}
\email{philsang.yoo@snu.ac.kr}

\title{3-dimensional mirror symmetry}
\newif\ifnotices
\newif\ifarxiv
\arxivtrue
\makeatletter
\renewcommand*\FXLayoutInline[3]{%
  {\@fxuseface{inline}
  
  \ignorespaces\noindent \ovalbox{\hspace{.01\textwidth} \begin{minipage}{.45\textwidth}
  	#3 \fxnotename{#1}: #2
  \end{minipage}\hspace{.01\textwidth}}}}
\makeatother
\usepackage{hyperref}
\hypersetup{
final,
linktoc=all
}

\begin{document}
\begin{abstract}
    This expository article discusses recent advances in understanding 3-dimensional mirror symmetry and the mathematical definitions of the Higgs and Coulomb branches.  This is a slightly expanded version of an article appearing in the Notices of the AMS.  
\end{abstract}
	\maketitle
\newcommand{\Sec}{\operatorname{Sec}}
\newcommand{\C}{\mathbb{C}}
\newcommand{\R}{\mathbb{R}}
\newcommand{\Z}{\mathbb{Z}}
\newcommand{\spinor}{\Sigma}
\newcommand{\spnorial}{U}
\newcommand{\alg}{\mathcal{A}}
\newcommand{\Spec}{\operatorname{Spec}}
\newcommand{\Hom}{\operatorname{Hom}}
\newcommand{\Bv}{\mathbf{v}}
\newcommand{\Bw}{\mathbf{w}}
\newcommand{\maybelater}[1]{}
\newcommand{\Sym}{\operatorname{Sym}}
\newcommand{\obs}{\operatorname{Obs}}
\newcommand{\NO}{N_{\mathcal{O}}}
\newcommand{\NK}{N_{\mathcal{K}}}
\newcommand{\GO}{G_{\mathcal{O}}}
\newcommand{\GK}{G_{\mathcal{K}}}
\newcommand{\XO}{X_{\mathcal{O}}}
\newcommand{\XK}{X_{\mathcal{K}}}
\newcommand{\XA}{X_n^{(\mathrm{A})}}
\newcommand{\XB}{X_n^{(\mathrm{B})}}
\newcommand{\XS}{X_n^{(*)}}
\newcommand{\fM}{\mathfrak{M}}
\newcommand{\disk}{\mathbb{D}}
\newcommand{\sPoin}{\mathfrak{a}}
\newcommand{\pdisk}{\mathbb{D}^{\circ}}
\newcommand{\arXiv}[1]{\href{http://arxiv.org/abs/#1}{\tt arXiv:\nolinkurl{#1}}}
\newcommand{\rkone}{M_{n\times n}^{\mathrm{rk}\leq 1}(\C)}
\newcommand{\notices}[1]{\ifnotices #1 \fi}
\newcommand{\arxiv}[1]{\ifarxiv #1 \fi}
\newcommand{\phil}[1]{\fxnote[inline,author=Phil]{\textcolor{red!50}{ #1}}}
\newcommand{\ben}[1]{\fxnote[inline,author=Ben]{\textcolor{blue!50}{ #1}}}

\newtheorem{itheorem}{Theorem}
\newtheorem{theorem}{Theorem}[section]
\newtheorem{proposition}[theorem]{Proposition}
\newtheorem{corollary}[theorem]{Corollary}
\newtheorem{lemma}[theorem]{Lemma}
\newtheorem{assumption}[theorem]{Assumption}
\newtheorem{definition}[theorem]{Definition}
\newtheorem{ex}[theorem]{Example}
\newtheorem{note}[theorem]{Note}
\newtheorem{ques}[theorem]{Question}
\newtheorem{conjecture}[theorem]{Conjecture}
\newtheorem{physics}[theorem]{Physics Motivation}

\listoffixmes

\section{Introduction}

\subsection{The House of Symplectic Singularities}
\label{sec:house}

Some have compared research in mathematics to searching through a dark room for a light switch\footnote{
	{\em Perhaps I could best describe my experience of doing mathematics in terms of entering a dark mansion. One goes into the first room, and it’s dark, completely dark. One stumbles around bumping into the furniture, and gradually, you learn where each piece of furniture is, and finally, after six months or so, you find the light switch. You turn it on, and suddenly, it’s all illuminated and you can see exactly where you were.} -- Andrew Wiles}. In other circumstances, it can be like walking through the same house during the day -- one can see all the furniture, but can still look through the drawers and cupboards for smaller nuggets of treasure. As enjoyable as such a treasure hunt is (and easier on the shins), discovering new rooms we haven't seen before may lead to even greater rewards. In some fields, this is just a matter of walking down the hall; the hard part is simply knowing which door to open. But even more exciting is finding a secret passage between two rooms we already thought we knew.  
	
	Of course, if you are not playing a game of {\it Clue}, secret passages can be hard to find.  You cannot just go tearing out walls and expecting them to be there.  However, in the late 20th and early 21st century, mathematicians found one remarkable source of such secret passages: quantum field theory (QFT).  
	
	What are called ``dualities'' in QFT often provide connections between mathematical objects that were totally unexpected beforehand.  For example, {\bf (2-dimensional) mirror symmetry} has shown that algebraic and symplectic geometers were actually living in the same house, though the passage between them is still quite poorly lit and harder to traverse than we would like.  
Unfortunately, employing these dualities in mathematics is not just a matter of bringing in a physicist with their x-ray specs; it is more like receiving an incomplete and weather-worn set of blueprints, possibly written in an unknown language, that hint at the right place to look.  Still, we get some very interesting hints.

For representation theorists, the most splendid and best explored of all mansions is the house of simple Lie algebras; while it is more than a century old, it still has many nooks and crannies with fascinating surprises.  It also has a rather innocent-looking little pass-through between rooms, called Langlands duality.  After all,  it is just transposing the Cartan matrix; most of us cannot keep the Cartan matrix straight from its transpose without looking it up anyway. The Langlands program has revealed the incredible depths of this simple operation.

 Many new wings have been found to this manor: Lie superalgebras, representations of algebraic groups in characteristic $p$, quiver representations, quantum groups, categorification, etc. Despite their diversity, they all rely on the same underlying framework of Dynkin diagrams. But in recent years, researchers have found a new extension more analogous to the discovery of many new series of Dynkin diagrams: the world of {\bf symplectic resolutions} and {\bf symplectic singularities}. According to an oft-repeated {\it bon mot}, usually attributed to Okounkov: ``symplectic singularities are the Lie algebras of the 21st century.'' 
 
 Interesting results about this particular annex started appearing around the turn of the 21st century, based on work of Kaledin, Bezrukavnikov and others.  Some time in 2007, my\footnote{All pronouns in this section are from the perspective of BW.}  collaborators Tom Braden, Nick Proudfoot, Tony Licata, and I noticed hints of another secret passage, connecting pairs of rooms (i.e. symplectic resolutions) there. Many coincidences were needed for the different rooms to line up precisely, making space for a secret passage. However, we were not able to step into the passage itself. Nevertheless, we found one very intriguing example: the secret passages we were looking for would generalize Langlands duality to many new examples.

Of course, you can guess from the earlier discussion what happened.  After I gave a talk at the Institute for Advanced Study in 2008, Sergei Gukov pointed out to me that physicists already knew that these secret passages should exist based on a known duality: {\bf 3-dimensional mirror symmetry}. As explained above, this definitely did not resolve all of our questions; to this day, an explanation of several of the observations we had made remains elusive.  More generally, 
this duality was poorly understood by physicists at the time (and many questions remain), but at least it provided an explanation of why such a passage should exist and a basis to search for it.  

In the 15 years since that conversation, enormous progress has been made on the connections between mathematics and 3-dimensional QFT.  The purpose of this article is to give a short explanation of this progress and some of the QFT behind it for mathematicians. It is, of necessity, painfully incomplete, but we hope that it will be a useful guide for mathematicians of all ages to learn more.  

\subsection{Plan of the paper}

Let us now discuss our plan with a bit more precise language.  A {\bf symplectic resolution} is a pair consisting of
\begin{enumerate}
    \item a singular affine variety $X_0$; and
    \item a smooth variety $X$ with an algebraic symplectic form which resolves the singularities of $X_0$. 
\end{enumerate}
The singular affine variety $X_0$ is a special case of a {\bf symplectic singularity}, which is a singular affine variety where the smooth locus is equipped with a symplectic form that is well-behaved at singularities.

The most famous example of a symplectic resolution is the {\bf Springer resolution}, where $X_0$ is the variety of nilpotent elements in a semi-simple Lie algebra $\mathfrak{g}$, and $X$ is the cotangent bundle of the flag variety of $\mathfrak{g}$.   You can reconstruct $\mathfrak{g}$ from the geometry of this resolution. Thus, one perspective on the house of simple Lie algebras is that the Springer resolution is really the fundamental object in each room of a simple Lie algebra, with all other aspects of Lie theory determined by looking at the Springer resolution from various different angles. 

Thus, simple Lie algebras lie at one end of a hallway, with many other doors that lead to other symplectic resolutions and singularities.  This leads to the natural question of whether any given notion for Lie algebras generalizes to other symplectic resolutions if we treat them like the Springer resolution of a new Lie algebra that we have never encountered.  For example, each symplectic singularity has a ``universal enveloping algebra'' which generalizes the universal enveloping algebra of a Lie algebra.    

 Two examples accessible to most mathematicians are:
\begin{itemize}
    \item The cotangent bundle $\XA=T^*\mathbb{CP}^{n-1}$ of complex projective space.     This can be written as 
    \begin{equation*}
    T^*\mathbb{CP}^{n-1} =\{(\ell, \phi)\in \mathbb{CP}^{n-1}\times M_{n\times n}(\C) \mid \phi(\C^n)\subset \ell, \phi(\ell)=\{0\}\}
    \end{equation*}
Projection to the second component is a resolution of $\rkone$, the space of $n\times n$ matrices of rank $\leq 1$.  
This cotangent bundle has a canonical symplectic form, which makes this resolution symplectic.
\item The cyclic group $\Z/n\Z$ acts on $\C^2$, preserving its canonical symplectic form, 
by the matrices
\[k\mapsto \begin{bmatrix} \exp(2\pi ik/n) & 0\\0& \exp(-2\pi ik/n)\end{bmatrix}
\]
The quotient $\C^2/(\Z/n\Z)$ has a unique symplectic resolution $\XB$ whose exceptional fiber is a union of $n-1$ copies of $\mathbb{CP}^1$'s that form a chain.  
\end{itemize}
We have an isomorphism $X_2^{(\mathrm{A})}\cong X_2^{(\mathrm{B})}$, but for $n>2$, these varieties have different dimensions.  
There are some intriguing commonalities when we look at certain combinatorial information coming out of these varieties.  Central to this are two geometric objects:
\begin{itemize}
    \item  The action of a maximal torus $T^{(*)}$ on $\XS$ for $*\in \{\mathrm{A},\mathrm{B}\}$ which preserves the symplectic structure.  One obvious invariant is the set of its fixed points of this torus.\footnote{$T^{(\mathrm{A})}$ is the diagonal matrices in $PGL_n(\C)$; $T^{(\mathrm{B})}$ the diagonal matrices in $SL_2(\C)$ modulo the $n$ torsion.}
    \item The affine variety $X_0$ has a unique minimal decomposition into finitely many smooth pieces with induced symplectic structures, generalizing the decomposition of nilpotent matrices into Jordan type.  
\end{itemize}
There are some intriguing coincidences between this pair of varieties:
\begin{enumerate}
    \item  We have isomorphisms \[\mathfrak{t}^{(\mathrm{A})}\cong H^2(\XB)\qquad \mathfrak{t}^{(\mathrm{B})}\cong H^2(\XA).\]
    We can make this stronger by noting that we match geometrically defined hyperplane arrangements on these spaces.\footnote{In $\mathfrak{t}$, the vectors where the vanishing set of the corresponding vector field jumps in dimension; in $ H^2(\XB)$, the Mori walls that cut out the ample cones of the different crepant resolutions of the same affine variety.}
    \item Both torus actions have the same number of fixed points, which is $n$; this also shows that the sum of the Betti numbers of $\XS$ is $n$.
    \item The stratifications on $X_0^{(\mathrm{A})}$ and  $X_0^{(\mathrm{B})}$ have the same number of pieces, which is $2$.\footnote{The smooth locus is one stratum, and in both cases, the other one is a single point.}
\end{enumerate}
It would be easy to dismiss these as not terribly significant, but they are numerical manifestations of a richer phenomenon. That is,
\begin{enumerate}
    \item[4.] the ``universal enveloping algebra'' of $\XS$  has a special category of representations that we call ``category $\mathcal{O}$'' (see \cite[\S 3]{BLPWgco}) and the categories $\mathcal{O}$ of $\XA$ and $\XB$ are {\bf Koszul dual}; the homomorphisms between projective modules in one category describe the extensions between simple modules in the other.
\end{enumerate} 
The other reason that we should not dismiss these ``coincidences'' is that the same statements 1.--4. apply to many pairs of symplectic singularities, which are discussed in \cite[\S 9]{BLPWgco}. These include all finite and affine type A quiver varieties and smooth hypertoric varieties. Some examples are self-dual:
\begin{itemize}
    \item $Y_n^{(\mathrm{A})}=Y_n^{(\mathrm{B})}=T^*\operatorname{Fl}_n$, the cotangent bundle of the variety of complete flags in $\C^n$.
    \item $Z_n^{(\mathrm{A})}=Z_n^{(\mathrm{B})}=\operatorname{Hilb}^{n}(\C^2)$, the Hilbert scheme of $n$ points in $\C^2$.  
\end{itemize}
 After suitable modification\footnote{In this case, the strata are the adjoint orbits of nilpotent elements, and the number of these is different for types $B_n$ and $C_n$.  We can recover a bijection by only considering {\bf special orbits}, of which there are the same number.} of 3., it also includes the Springer resolutions of Langlands dual pairs of Lie algebras.
 
 This mysterious duality on the set of symplectic singularities and their resolutions has obtained the name of ``symplectic duality'' for its connection of two apparently unrelated symplectic varieties.
 
\begin{ques}\label{ques:duality}
Is there an underlying principle that explains statements 1.-4., that is, which explains the symplectic duality between these pairs of varieties?
\end{ques}

As discussed above, work on QFT in dimension 3 suggests that the answer to this question is closer to ``yes'' than it is to ``no.''  Our aim in this article is to explain the basics of why this is so and what it tells us about mathematics.  

We can break this down into two sub-questions: 
\begin{enumerate}
	\item[Q1.] What are 3d $\mathcal{N}=4$ SUSY QFTs and their topological twists?
	\item[Q2.] What do they have to do with symplectic duality?
\end{enumerate}
In Section \ref{sec:physical-origin}, we will provide an answer to the questions, which we now briefly summarize.

First, every 3-dimensional topological quantum field theory (TQFT) gives us a Poisson algebra (see \cite{BBBDN} for more discussion of this construction).  In many cases, this ring is the coordinate ring of a symplectic singularity $X_0$, and all the examples discussed above can be constructed in this way.  
Given a QFT, a choice of a topological twist gives rise to a TQFT. In fact, for a 3d $\mathcal{N}=4$ theory $\mathcal{T}$, there are two such choices, called the $A$-twist and the $B$-twist. Hence each 3d theory gives two symplectic singularities $\fM_A(\mathcal{T})$ and $\fM_B(\mathcal{T})$ called the {\bf Coulomb branch} and {\bf Higgs branch} of the theory.

The pairs of symplectic varieties $X^{(\mathrm{A})}$ and $X^{(\mathrm{B})}$  (similarly, $Y,Z$, etc.) all turn out to be the Coulomb and Higgs branches of a single theory $\mathcal{T}$. Then the statements 1.-4. can be understood in terms of the physical duality referenced in Section \ref{sec:house}, called ``3-dimensional mirror symmetry.''

\notices{This is a very large topic, and due to constraints on the length and number of references,  we will concentrate on the relationship to symplectic resolutions of singularities, giving relatively short shrift to the long and rich literature in physics on the topic;  the introduction of \cite{bullimoreBoundariesMirror2016}  will lead the reader to the relevant references, starting from the original work of Intrilligator--Seiberg and Hanany--Witten, which laid the cornerstone of this theory.}

\arxiv{
This is a very large topic, and due to constraints on the length and number of references,  we will concentrate on the relationship to symplectic resolutions of singularities \cite{beauvilleSymplecticSingularities2000}.  In particular, we will give relatively short shrift to the long and rich literature in physics on the topic;  the introduction of \cite{bullimoreBoundariesMirror2016}  will lead the reader to the relevant references, starting from the original work of Intrilligator--Seiberg   \cite{intriligatorMirrorSymmetry1996} and Hanany--Witten \cite{hananyTypeIIB1997} which laid the cornerstone of this theory.
}

Just as the 2-dimensional mirror symmetry known to mathematicians suggests that complex manifolds and symplectic manifolds (with extra structure) come in pairs whose relationship is hard to initially spot, 3-dimensional mirror symmetry rephrases our answer to Question \ref{ques:duality}: the Coulomb branch of one theory can also be thought of as the Higgs branch of its dual theory: $\fM_A(\mathcal{T})=\fM_B(\mathcal{T}^{\vee})$.  Thus, we can also describe our dual pairs of symplectic varieties as the Higgs branches of dual theories $(\fM_B(\mathcal{T}), \fM_B(\mathcal{T}^{\vee}) )$.

This answer is not as complete as we would like, since we cannot construct 3-dimensional QFTs as rigorous mathematical objects. We can only work with mathematical rigor on certain aspects of some classes of theories, the most important of which are {\bf linear gauge theories}.  In these cases, we have mathematical definitions of the Higgs and Coulomb branches and thus can prove mathematical results about them. 

 In Section \ref{sec:Higgs-and-Coulomb}, we will review these constructions of the Higgs and Coulomb branches in the case of linear gauge theories. The former of these constructions has been known to mathematicians for many decades \cite{hitchinHyperkahlerMetrics1987}, but the construction of Coulomb branches was a surprise even to physicists when it appeared in 2015 \cite{BFN}, and is key to the progress we have made since that time. 
 
 These varieties are the keystones of a rapidly developing research area that combines mathematics and physics. In particular, they point the way to understanding a mirror symmetry of 3-dimensional theories that is not only a counterpart to the mirror symmetry known to mathematicians (which is 2-dimensional mirror symmetry)  but also provides an enrichment of the geometric Langlands program (which comes from a duality of 4-dimensional theories). 
 
 We will conclude the article in Section \ref{sec:advanced} with a brief discussion of interesting directions of current and future research to give the interested reader guidance on where to turn next.

\section*{Acknowledgements}

The colleagues and collaborators who have helped us to learn this material, including by giving feedback on this article, are almost too numerous to list.  Justin Hilburn deserves particular appreciation for many conversations and insights on these topics, as do Tudor Dimofte, Davide Gaiotto, Kevin Costello, David Ben-Zvi, Nick Proudfoot, Tom Braden, Anthony Licata, Contantin Teleman, Andrei Okounkov,  Mina Aganagi\'c, Richard Rimany\'i, Hiraku Nakajima and  Sam Raskin.  

This research was also supported by Perimeter Institute for Theoretical Physics, through the 6 years that B.W. has spent there and many visits by P.Y. as well as two ``QFT for Mathematicians'' conferences that were key to the development of this paper. Research at Perimeter Institute is supported by the Government of Canada through the Department of Innovation, Science and Economic Development and by the Province of Ontario through the Ministry of Research and Innovation.  This work was supported  by the Discovery Grant RGPIN-2018-03974 from the Natural Sciences and Engineering Research Council of Canada, New Faculty Startup Fund from Seoul National University, and the National Research Foundation of Korea(NRF) grant funded by the Korea government(MSIT) No. 2022R1F1A107114212.

\section{Physical Origin}
\label{sec:physical-origin}

\subsection{QFT}
\label{sec:QFT}

In this section, we will give a very short introduction to (Euclidean) QFT.
Typically, a QFT has the following input data:
\begin{enumerate}
	\item (\textbf{spacetime}) a $d$-dimensional Riemannian manifold $(M,g)$;
	\item (\textbf{fields}) a fiber bundle $B$ over $M$ and the space $\mathcal F=\mathcal{F}(M)=\Gamma(M,B)$ of sections of $B$ over $M$;
	\item (\textbf{action functional}) a functional $S \colon\mathcal F \to \R$;
\end{enumerate}
In very rough terms, $\mathcal F$ should be viewed as the space of all possible states of a physical system, while the function $S$ controls which states will likely be physically achieved. 
\maybelater{as it would contribute most to the integral $ \int_{ \phi \in \mathcal{F}} e^{ - S(\phi)/\hbar } D\phi $ by the method of steepest descent.}

In a classical physical system, we want to think about measuring quantities, such as the velocity or position of a particle.  We can formalize this in the notion of an \textbf{observable}, which is, by definition, a functional $O \colon \mathcal{F}\to \R$.
A particularly important type is \textbf{local operators} at $x$ that depend only on the value of a field or its derivatives at $x$. 


\begin{ex}[Free scalar field theory]\hfill 
\begin{enumerate}
	\item a (compact) Riemannian manifold $(M,g)$;
	\item $B=M\times \R$ so that $\mathcal{F}=C^\infty(M)$;
	\item $S \colon\mathcal C^\infty(M) \to \R$ given by $S(\phi)=\int_M  \phi\Delta_g \phi \mathrm{Vol}_g$ where $\Delta_g$ is the Laplacian of the metric $g$ and $\mathrm{Vol}_g$ is the volume form associated to $g$.
\end{enumerate}
In the case of $M=\R$, for any point $x\in \R$, the functionals $O_x, O_x^{(1)} \colon C^\infty(\R) \to \R$ defined by $O_x(\phi)= \phi(x)$ and $O_x^{(1)}(\phi) =  \phi'(x) $ are local operators at $x$.
\end{ex}

Two other types of field theories play an important role for us:
	\begin{enumerate}
		\item Let $G_c$ be a compact Lie group. When $\mathcal{F}$ consists of connections on a principal $G_c$-bundle over $M$, such a field theory is called a \textbf{gauge theory} and $G_c$ is called the \textbf{gauge group} of the theory.
		\item Let $X$ be a manifold. When $\mathcal{F}$ consists of maps from $M$ to $X$, such a theory is called a \textbf{$\sigma$-model} and $X$ is called the \textbf{target} of the $\sigma$-model.  In this case, $B=M\times X$.  
	\end{enumerate}

One insight of the quantum revolution in physics is that a physical system cannot be described by a single field, which would have a well-defined value for each observable.  Instead, we can only find the expectation values of observables as integrals, where a measure depending on the action accounts for how probable states are. These integrals are often written notionally in the form 
\begin{equation}
    \langle O\rangle :=  \int_{\phi\in \mathcal{F}(M)} e^{ - S(\phi)/ \hbar  } O(\phi) D \phi.  \label{eq:expected-value}\notag
\end{equation} However, in many cases, these integrals do not make sense because the space $\mathcal{F}(M)$ is often infinite-dimensional, and as a result, the Lebesgue measure $D\phi$ cannot be defined.

\notices{More generally, given observables $O_i$'s which only depend on the values of the fields on open sets that do not overlap, we consider the integrals of the following form
\[ \langle O_1,\cdots,O_n\rangle :=  \int_{\phi\in \mathcal{F}(M)} e^{ - S(\phi)/ \hbar  } O_1(\phi)\cdots O_n(\phi) D \phi.   \]	
These are called the \textbf{correlation functions} of the theory and the main objects of study in a QFT. One may also understand the integral as the correlation function of a single observable, as the notion of \textbf{operator product} allows one to express products of $O_1,\cdots,O_n$ as a single observable.}

\arxiv{On the other hand, note that one can define the space of fields $\mathcal{F}(U)=\Gamma(U,B)$ over any open subset $U$ of $M$ and hence observables as well by $\obs(U) = \operatorname{Fun}(\mathcal{F}(U))$.\footnote{One has $\obs(U)\to \obs(V)$ for any $U\subset V$ and the space $\obs(x)$ of local operators at $x$ then may be identified as the inverse limit $\displaystyle\varprojlim_{x \in U} \obs(U)$ over all open neighborhoods $U$ of $x$.} In quantum theory, an observation itself disturbs the system so one is not able to make two observations at the same point of the spacetime $M$ in a coherent way. However, we still have a way to combine those at different points, that is, for disjoint open sets $U_1$ and $U_2$ inside $V$, we expect to have a map $\obs(U_1) \otimes \obs (U_2) \to \obs (V)$ that captures an \textbf{operator product}:
\begin{equation*}
\tikz{
	\node[draw, circle, inner sep=0.5pt ] (a) at (-3.9,0) {$U_1$};
	\node[draw, circle, inner sep=2pt] (b) at (-3.2,0) {$U_2$};	
	\node[draw, circle, inner sep=15pt] (c) at (-3.5,0) {};	
	\node at (-3.5,0.5) {$V$}
}
\end{equation*}
More generally, given observables $O_i$ that depend only on the values of the fields on open sets that don't overlap, we consider the integrals of the following form
\[ \langle O_1,\cdots,O_n\rangle :=  \int_{\phi\in \mathcal{F}(M)} e^{ - S(\phi)/ \hbar  } O_1(\phi)\cdots O_n(\phi) D \phi.   \]	
These are called the \textbf{correlation functions} of the theory and the main objects of study in a QFT.} 

\subsection{TQFT}
\label{sec:TQFT}

In the framework of Atiyah and Segal, a $d$-dimensional \textbf{topological quantum field theory\arxiv{\footnote{For physicists, TQFT usually means a QFT where correlation functions are independent of continuous change of the metric $g$ of the spacetime $M$. To distinguish these two related but distinct notions, we use the term ``functorial TQFT'' for a TQFT in the sense of Atiyah and Segal. At the level of ideas, whenever one has a TQFT in the sense of physicists, one may imagine having a functorial TQFT, although it may be daunting to actually construct one.}} (TQFT)} is a symmetric monoidal functor $Z$ from the category $(\operatorname{Bord}_d,\amalg,\emptyset)$  to the category $(\operatorname{Vect}_{\mathbb{C}},\otimes,\C)$ of complex vector spaces. Objects of $\operatorname{Bord}_d$ are closed oriented $(d-1)$-manifolds $N$, a morphism from $N$ to $N'$ is a diffeomorphism class of a $d$-dimensional bordism $M$ from $N$ to $N'$, and the monoidal structure is given by disjoint union $\amalg$ with the empty set $\emptyset$ being the unit object.

Regarding a closed $d$-manifold $M$ as a bordism from $\emptyset$ to $\emptyset$ yields \arxiv{a linear map $Z(M) \colon \C  \to \C $, or }a complex number $Z(M)$. Physically, one should imagine that $Z(M)=\int_{\phi \in \mathcal{F}(M)} e^{ - S(\phi)/ \hbar  } D \phi $. On the other hand, the complex vector space $Z(N)$ attached to a closed $(d-1)$-manifold $N$ is the Hilbert space of states on $N$\arxiv{ of the physical system described by the TQFT.} 

\arxiv{For example, first, consider the cases where $d=1$: we obtain a system called ``topological quantum mechanics,''  where $H=Z(\operatorname{pt})$ is the space of states of a single particle of the system and time evolution is trivial due to the topological nature of the theory. Under the state-operator correspondence, we have $Z(S^0)\cong \operatorname{End}(H)$, which we can identify with $H\hat{\otimes}H$ via the action $(a\otimes b)v=\langle b,v\rangle a$;  physicists will prefer writing $|a\rangle\langle b| $ in place of $a\otimes b$ to suggest this multiplication.} 

Suppose $d=2$. Since any closed oriented 1-manifold is a disjoint union of copies of circles, it is enough to describe $Z(S^1)$. Moreover, the map associated to a pair of pants yields a linear map $m\colon Z(S^1)\otimes Z(S^1)\to Z(S^1)$ and the one associated to a disk is a linear map $ u \colon  Z(\emptyset)\to Z(S^1)$:
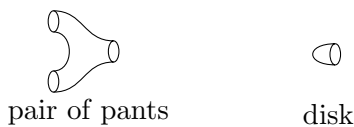
\begin{center}
\begin{tikzpicture}[tqft/flow=east]
\begin{scope}[scale=0.4,every tqft/.style={transform shape},tqft/boundary lower style={draw}]
\node[tqft/reverse pair of pants, draw]   (a) {};
\path (a.outgoing boundary 1) +(-1,-2) node[font=\small] {\(\text{ pair of pants}\)};
\end{scope}
\end{tikzpicture}\hspace{3em} 
\begin{tikzpicture}[tqft/flow=east]
\begin{scope}[scale=0.4,every tqft/.style={transform shape},tqft/boundary lower style={draw}]
\node[tqft/cap, draw]   (a) {};
\path (a.outgoing boundary 1) +(-0.4,-2) node[font=\small] {\(\text{ disk}\)};
\end{scope}
\end{tikzpicture} \vspace{-0.5em}
\end{center}
\arxiv{In fact, the following pictures ensure that $m$ and $u$ define a unital associative algebra structure on $Z(S^1)$:
\begin{center}
\begin{tikzpicture}[tqft/flow=east]
\begin{scope}[scale=1/4,every tqft/.style={transform shape},tqft/boundary lower style={draw}]
 \node[tqft/reverse pair of pants, draw, anchor=outgoing boundary 1] (a) {};
 \node[tqft/reverse pair of pants, draw, anchor=outgoing boundary 1] at (a.incoming boundary 1) (b) {};
 \node[tqft/cylinder to prior, draw, anchor=outgoing boundary 1]  at (a.incoming boundary 2) (c) {};
\path (a.outgoing boundary 1) +(1.5,0) node {\(=\)};
\path (a.outgoing boundary 1) +(1.5,-3.5) node[font=\small] {\(\text{associativity}\)};
\path (a.outgoing boundary 1) +(3,0) node[tqft/reverse pair of pants,draw, anchor=incoming boundary 1] (d) {};
 \node[tqft/reverse pair of pants, draw, anchor=incoming boundary 2]  at (d.outgoing boundary 1) (e) {};
 \node[tqft/cylinder to next, draw, anchor=outgoing boundary 1]  at (e.incoming boundary 1) (f) {};
\end{scope}
\end{tikzpicture}
\hspace{2em}  \begin{tikzpicture}[tqft/flow=east]
\begin{scope}[scale=1/4,every tqft/.style={transform shape},tqft/boundary lower style={draw}]
 \node[tqft/reverse pair of pants, draw, anchor=outgoing boundary 1]  (a) {};
 \node[tqft/cap, draw, anchor=outgoing boundary 1] at (a.incoming boundary 1)  (b) {};
 \node[tqft/cylinder, draw, anchor=outgoing boundary 1]  at (a.incoming boundary 2) (c) {};
\path (a.outgoing boundary 1) +(1.5,0) node  {\(=\)};
\path (a.outgoing boundary 1) +(4,-3.5) node[font=\small] {\(\text{unit axiom}\)};
\path (a.outgoing boundary 1) +(3,0) node[tqft/cylinder,draw, anchor=incoming boundary 1] (d) {};
\path (a.outgoing boundary 1) +(6.5,0) node  {\(=\)};
 \path (d.outgoing boundary 1) +(7,0) node[tqft/reverse pair of pants, draw, anchor=outgoing boundary 1] (e) {};
 \node[tqft/cap, draw, anchor=outgoing boundary 1] at (e.incoming boundary 2)  (f) {};
 \node[tqft/cylinder, draw, anchor=outgoing boundary 1]  at (e.incoming boundary 1) (g) {};
\end{scope}
\end{tikzpicture} \vspace{-0.5em}
\end{center}
Moreover, the multiplication is commutative because one can switch the order of multiplication. The algebra $Z(S^1)$ can be equipped with the structure of a commutative Frobenius algebra by also considering the diagrams above read right-to-left. In fact, $d=2$ TQFTs are classified by commutative Frobenius algebras.}
\notices{Topological arguments show that these maps and others from the reversed picture induce a commutative Frobenius algebra structure on $Z(S^1)$.}

Note that one can apply a similar idea to any $d$-dimensional TQFT $Z$ to show that $Z(S^{d-1})$ obtains a commutative algebra structure for $d\geq 2$ using the cobordism where we remove two disjoint $d$-balls from the interior of a $d$-ball. 
\notices{When we interpret $Z(S^{d-1})$ as the space of local operators of the theory, this product has a physical meaning:  it is precisely the operator product introduced above.\footnote{An important warning for the reader: we will considering topological twists of QFTs below, which do not always produce TQFTs in the framework above, since the maps defined by some cobordisms may not converge. For example, $Z(S^{d-1})$ will not be finite-dimensional in the examples we consider.}

Since this is a commutative $\C$-algebra, $Z(S^{d-1})$ can be interpreted as the coordinate ring of an algebraic variety. In fact, the spectrum $\mathfrak{M}=\Spec Z(S^{d-1})$ has a physical interpretation as well:  it is the moduli space of vacua of the theory.  This reflects the fact that at a vacuum state, which by definition is a linear map $\langle-\rangle \colon Z(S^{d-1})\to\C$, measurements at distant points cannot interfere so that $\langle O_1 O_2\rangle =\langle O_1\rangle \langle O_2\rangle$. Thus, $O \mapsto  \langle O\rangle$ defines a ring map $Z(S^{d-1})\to \C$, and a point in the spectrum.}

\arxiv{One may ask, where does this structure come from, physically speaking? That is, while we know that $Z(S^{d-1})$ is the space of states on $S^{d-1}$, why should a Hilbert space be equipped with a multiplication at all?

The idea has to do with the notion of local operators of QFT. Since the theory is assumed to be topological, the space of local operators at $x$ may well be faithfully represented by $\obs(D^d)$ for a $d$-dimensional disk $D^d$. As a disk $D^d$ has $S^{d-1}$ as its boundary, one should imagine the picture of a disk above as providing a linear map $\obs(D^d)\to Z(S^{d-1})$ given by $O\mapsto O|0\rangle$, where $|0\rangle$ is the vacuum state of the Hilbert space $Z(S^{d-1})$, a local operator $O$ is placed at a point on $D^d$, and then $O|0\rangle$ reads off the resulting state of the operator.

A non-trivial claim that holds in the context of TQFT is that the linear map $\obs(D^d)\to Z(S^{d-1})$ is an isomorphism and hence one can take $Z(S^{d-1})$ to be the space of local operators (at any point $x$) of a given theory. This claim is known as the \textbf{state-operator correspondence}. This then solves the riddle; the commutative algebra structure on $Z(S^{d-1})$ is the operator product of local operators of TQFT. It is also clear that the picture of a pair of pants (generalized using $S^{d-1}$ instead of $S^1$) can also be regarded as the picture of two disjoint disks $D^d$ inside a bigger disk $D^d$, that is, the one of an operator product above.}

In many examples of applications of the idea of physics to mathematics, the perspective of TQFT provides a useful guiding principle. Before discussing how to use the idea, let us explain how one may obtain a TQFT starting from a QFT.

\subsection{From QFT to TQFT}

There are two well-known ways to construct a TQFT, that is, a theory which is independent of a metric of the spacetime manifold. One is to begin with a space of fields and action functional which do not depend on a metric. For example, Chern--Simons theory is one such theory. This approach is quite limited and leads to relatively few examples.  Many more examples arise from applying a {\bf topological twist} to a supersymmetric field theory (which depends on a metric). Let us briefly review the latter idea.

Consider $M=\R^d$ with the standard metric. In this case, the isometry group $\operatorname{ISO}(d,\R)=\operatorname{SO}(d,\R)\ltimes \R^d$ is called the \textbf{Poincar\'e group} and acts on $\R^d$ by rotation and translation. We will only consider field theories on $\R^d$ where the action functional is equivariant under the induced action on the space of fields. 

We will also only consider theories where the space of sections $\mathcal{F}$ is $\Z/2\Z$-graded; this arises physically from the spin angular momentum of particles, and thus the natural classifications of particles into bosons (even) and fermions (odd).
We call a field theory {\bf supersymmetric (SUSY)} if it admits non-trivial ``odd symmetries'', which one calls {\bf supercharges}. 

More precisely, this means that the space $\mathcal{F}$ carries an action of a Lie superalgebra $\sPoin$ called a \textbf{super-Poincar\'e algebra} whose even part is the Poincar\'e algebra  $\sPoin_0=\mathfrak{so}(d)\ltimes \R^d$ and whose odd part $\sPoin_1=\Sigma$ consists of copies of spin representations of $\mathfrak{so}(d)$. 
A Lie bracket is given by the action of $\mathfrak{so}(d)$ on $\Sigma$, as well as a symmetric\footnote{In the world of super Lie algebras, Lie bracket is symmetric if both inputs are odd!} pairing $\Gamma \colon \Sigma\otimes \Sigma\to \R^d$  of $\mathfrak{so}(d)$-representations.

For simplicity, we work with a complexification of the supersymmetry algebra from now on; this is mostly harmless for the purpose of discussing twists.
\begin{ex}\hfill
\begin{enumerate}
	\item The $d=2$, $(\mathcal N_+,\mathcal N_-)$ supersymmetry algebra $\sPoin_{d=2}$ has odd part $ S_+\otimes W_+\oplus S_-\otimes W_-$, where $S_\pm \cong \C $ are the two spin representations of $\mathfrak{so}(2)$ and $\dim W_\pm= \mathcal N_\pm$. The pairing $\Gamma \colon \Sigma\otimes \Sigma\to \C^2$ is induced by the isomorphism $S_+^{\otimes 2} \oplus S_-^{\otimes 2} \cong \C^2$ as $\mathfrak{so}(2)$-representations.
	\item The $d=3$, $\mathcal N=k$ supersymmetry algebra $\sPoin_{d=3}$ has odd part $S \otimes W$, where $S\cong \C^2$ is the spin representation of $\mathfrak{so}(3)$ and $\dim W=\mathcal N$. The pairing $\Gamma \colon \Sigma\otimes \Sigma\to \C^3$ is induced by the isomorphism $ \operatorname{Sym}^2 S\cong \C^3$ as $\mathfrak{so}(3)$-representations.
\end{enumerate}	
\end{ex}

Finally, in order to extract a TQFT from a SUSY theory, suppose that one has chosen a supercharge $Q$ of a SUSY algebra such that $[Q,Q]=0$. Since $Q$ is odd, this means $\frac{1}{2}[Q,Q]=Q^2$ acts as zero in any representation of $\sPoin$. Hence, one can consider $\mathcal{F}$ or even $\sPoin$ itself as a $\Z/2\Z$ graded complex, and take its $Q$-cohomology. Necessarily, this procedure results in a simpler theory, which one calls a \textbf{twist} or a \textbf{twisting}.

If an element $x\in \R^d\subset \sPoin$ is in the image of $[Q,-]$, then translation by $x$ will be trivial in the twisted theory. \arxiv{In particular, if $O_i$'s are $Q$-closed local operators at different points $x_i$, then the correlation function $\langle O_1\dots O_n\rangle$ will be unchanged if one continuously moves the points $x_i$ along directions in the image of $[Q,-]$, as long as they don't collide.} The most important case for us is if the image of $Q$ fills in all of $\C^d$.  In this case, the dependence on position vanishes and the theory becomes topological; consequently, the twisted theory is called a \textbf{topological twist} of the original theory.

Whether a topological twist exists is purely dependent on the super-Poincar\'e algebra $\sPoin$, and thus on $d$ and $\mathcal{N}$.  \arxiv{We will use two important facts about this dependence on the number of supercharges.} Let $\mathfrak{a}_{d=3}$ (resp. $\mathfrak{a}_{d=2}$) be the supersymmetry algebra with $d=3$ (resp. $d=2$) and $\mathcal{N}=n$ (resp. $\mathcal{N}=(n_+,n_-)$) supersymmetry.  By a standard argument (see, e.g., \cite[\S\S 11.2 \& 12.1]{elliottTaxonomyTwists2022}), we have:
\begin{itemize}
	\item In the case $d=3$ (resp. $d=2$) there is a topological twist if and only if $n\geq 4$ (resp. $n_\pm\geq 2$).  
	\item In the case where $n= 4$ (resp. $n_\pm= 2$), there are exactly 2 topological twists up to appropriate symmetry, which we denote by $Q_{\operatorname{A}}$ and $Q_{\operatorname{B}}$.
\end{itemize}
\maybelater{It turns out that for $d=2$, $\mathcal{N}=(2,2)$ supersymmetry algebras, there are precisely 2 supercharges up to appropriate symmetry which realize topological twists, which one calls $Q_{\operatorname{A}}$ and $Q_{\operatorname{B}}$. Similarly, for $d=3$, that $\mathcal{N}=4$ supersymmetry algebras also have precisely 2 supercharges up to appropriate symmetry which realize topological twists, which one calls $Q_{\operatorname{A}}$ and $Q_{\operatorname{B}}$. }
\subsection{Mirror Symmetry}

When $X$ is a Calabi--Yau manifold,\arxiv{\footnote{The story can be extended to a K\"ahler manifold, but we won't for simplicity. For the same reason, we won't discuss the role of a superpotential both in 2d and 3d mirror symmetry.}} there is a physics construction of a 2-dimensional $\mathcal N=(2,2)$ SUSY $\sigma$-model $\mathcal{T}^{d=2}(X)$ with target $X$. If we twist with respect to $Q_{\operatorname{A/B}}$,  the resulting TQFT is called the A/B-model $\mathcal{T}^{d=2}_{\operatorname{A/B}}(X)$. The A model depends on the symplectic topology of $X$, and the B-model on the complex geometry of $X$.

There is a remarkable duality, called \textbf{mirror symmetry}, on the set of such SUSY $\sigma$-models, which identifies $\mathcal{T}^{d=2}(X)$ and $\mathcal{T}^{d=2}(X^\vee)$ for another {\bf mirror dual} Calabi--Yau manifold $X^\vee$. 
Moreover, this duality is compatible with topological twists: the identification of $\mathcal{T}^{d=2}(X)$ and $\mathcal{T}^{d=2}(X^\vee)$ is compatible with an involution of the $d=2$, $\mathcal N =(2,2)$ SUSY algebra which exchanges $Q_{\operatorname{A}}$ and $Q_{\operatorname{B}}$.  Therefore, the $d=2$ TQFTs $\mathcal{T}^{d=2}_{\operatorname{A}}(X)$ and $\mathcal{T}^{d=2}_{\operatorname{B}}(X^\vee)$ should be equivalent. This idea has resulted in several marvelous predictions. The most famous is that the numbers of rational curves of degree $d$ on a quintic 3-fold, understood as the correlation functions of $\mathcal{T}^{d=2}_{\operatorname{A}}(X)$, should be equal to the correlation functions of $\mathcal{T}^{d=2}_{\operatorname{B}}(X^\vee)$, which can be more easily computed.

The remarkable success of $d=2$ mirror symmetry \arxiv{(in mathematics simply ``mirror symmetry'')}motivates the consideration of an analogous duality, called \textbf{3-dimensional mirror symmetry}, for $d=3$, $\mathcal{N}=4$ SUSY field theories, which identifies two superficially different theories, say $\mathcal{T}$ and $\mathcal{T}^\vee$. Just as before, there are still two interesting topological twists $Q_{\operatorname{A}}$ and $Q_{\operatorname{B}}$ in the super-Poincar\'e algebra, and an automorphism of $\sPoin$ which switches these. By the same logic, we have an equivalence of topologically twisted theories between $\mathcal{T}_{\operatorname{A}}$ and $\mathcal{T}_{\operatorname{B}}^{\vee}$, which we write $Z_{\operatorname{A}}^{\mathcal{T}}$ and $Z_{\operatorname{B}}^{\mathcal{T}^\vee}$, respectively, to emphasize the TQFT perspective. \arxiv{This might seem like a pure abstraction, but just as in the $d=2$ case, we can derive more familiar mathematical objects from these topologically twisted theories.  }

We will focus on understanding the algebras $\alg_{\operatorname{A/B}}(\mathcal{T})=Z_{\operatorname{A/B}}^{\mathcal{T}}(S^2)$. As discussed in Section \ref{sec:TQFT}, the algebraic varieties \[\mathfrak{M}_{\operatorname{A}}(\mathcal{T})=\Spec Z_{\operatorname{A}}^{\mathcal{T}}(S^{2})\qquad \mathfrak{M}_{\operatorname{B}}(\mathcal{T})=\Spec Z_{\operatorname{B}}^{\mathcal{T}}(S^{2})\]
are the moduli spaces of vacua of the respective theories. We will call these the {\bf Coulomb branch} $\mathfrak{M}_{\operatorname{A}}(\mathcal{T})$ and {\bf Higgs branch}  $\mathfrak{M}_{\operatorname{B}}(\mathcal{T})$ of the theory $\mathcal{T}$\arxiv{; these objects are usually described a little differently in the physics literature, but for the theories of interest to us, these will be the same.} 
Of course, the identification of local operators $\alg_{\operatorname{A}}(\mathcal{T})$ in one theory with $\alg_{\operatorname{B}}(\mathcal{T}^{\vee})$ in the mirror theory is one of the most important features of mirror symmetry in this case as well:
\[\mathfrak{M}_{\operatorname{A}}(\mathcal{T})\cong \mathfrak{M}_{\operatorname{B}}(\mathcal{T}^{\vee})\qquad \qquad \mathfrak{M}_{\operatorname{B}}(\mathcal{T})\cong \mathfrak{M}_{\operatorname{A}}(\mathcal{T}^{\vee}).\] Thus we call the varieties $\mathfrak{M}_{\operatorname{A}}(\mathcal{T})$ and $ \mathfrak{M}_{\operatorname{B}}(\mathcal{T})$ {\bf mirror} to each other, or symplectic duals in the terminology of \cite{BLPWgco}. These varieties have the virtue of being familiar types of mathematical objects, while still carrying much of the structure of the theory $\mathcal{T}$.

\maybelater{
 An important feature of this duality is that different components of the moduli space of vacua, that is, the Coulomb and Higgs branches are swapped under duality.  To discuss the moduli of vacua, it is useful to explain what a vacuum state is. For a general QFT on $\R^d$, a state is what gives a number when paired with an observable on $\R^d$, namely, a linear map $\langle -\rangle \colon  \obs(\R^d)\to \C$. A state is called a \textbf{vacuum} if it is translation invariant and satisfies the \textbf{cluster decomposition property}; that is, for any observables $O_1,O_2\in \obs(U)$, the value $\langle O_1 \ast \tau_xO_2\rangle - \langle O_1\rangle \langle O_2\rangle$ tends to zero as $x$ approaches $\infty$, where $\tau_x$ is a translation of an observable by $x \in \R^d$ and we use an operator product $\ast$. This is supposed to capture the idea that observations at a vacuum cannot interfere with each other when they are far from each other.
 
Now, if a given theory is topological, then translation invariance is automatic and the cluster decomposition property amounts to having a ring homomorphism $\langle -\rangle \colon \obs(\R^d)\to \C$. Suppose that we are given a $d$-dimensional functorial TQFT $Z$. Then thanks to the identification between $\obs(\R^d)$ and $Z(S^{d-1})$ as commutative algebras, the moduli space of vacua of $Z$ should be given by $\Spec Z(S^{d-1})$.


To see the relevance of this observation to the situation at hand, we note that a 3d $\mathcal{N}=4$ theory also admits two topological twists by $Q_{\operatorname{A}}$ and $Q_{\operatorname{B}}$, say $\mathcal{T}_{\operatorname{A}}$ and $\mathcal{T}_{\operatorname{B}}$. Assuming that they define $d=3$ functorial TQFTs, say $Z_{\operatorname{A}}$ and $Z_{\operatorname{B}}$, one should be able to define moduli of vacua of $\mathcal{T}_{\operatorname{A}}$ and $\mathcal{T}_{\operatorname{B}}$ as $\Spec Z_{\operatorname{A}}(S^2)$ and $\Spec Z_{\operatorname{B}}(S^2)$, respectively. It turns out that the moduli spaces of vacua of $\mathcal{T}_{\operatorname{A}}$ and of $\mathcal{T}_{\operatorname{B}}$ are what physicists call the Coulomb branch and the Higgs branch of the original theory $\mathcal{T}$, respectively. Henceforth, we write $\mathfrak{M}_{\operatorname{A}}=\mathfrak{M}_{\operatorname{A}}(\mathcal{T})$ for the Coulomb branch of $\mathcal{T}$ and $\mathfrak{M}_{\operatorname{B}}=\mathfrak{M}_{\operatorname{B}}(\mathcal T)$ for the Higgs branch of $\mathcal{T}$.

3d mirror symmetry is also known to respect topological twists in that $\mathcal T_{\operatorname{A}}$ and $\mathcal{T}^\vee_{\operatorname{B}}$ are equivalent. In particular, their moduli of vacua should be the same and one should have $\mathfrak{M}_{\operatorname{A}}(\mathcal{T}) \cong \mathfrak{M}_{\operatorname{B}}(\mathcal{T}^\vee )$ and $\mathfrak{M}_{\operatorname{B}}(\mathcal{T}) \cong \mathfrak{M}_{\operatorname{A}}(\mathcal{T}^\vee )$.}

\maybelater{To extract any nontrivial content out of this, we need some examples of $d=3$, $\mathcal{N}=4$ theories. When $X$ is a hyperk\"ahler manifold, there is a procedure to define such a theory. Better yet, when a compact Lie group $G_c$ acts on $X$ in a hyper-Hamiltonian fashion, there is a way of constructing such a SUSY gauged $\sigma$-model, which we denote by $\mathcal{T}(X,G)$. Then physics suggest examples where $(X,G)$ is mirror dual to $(X^!,G^!)$\footnote{It is a duality of pairs; $X^!$ or $G^!$ alone doesn't make sense.}. In the following, we will discuss the mathematical constructions of $\mathfrak{M}_{\operatorname{A}}$ and $\mathfrak{M}_{\operatorname{B}}$ of $\mathcal{T}(X,G)$ and discuss their mirror symmetry statement.

\subsection{Future Direction}
\phil{this part is not thought out hard enough, and I would like to discuss what/how to write}
Before explaining the mathematical constructions of the moduli spaces of vacua of $\mathcal{T}(X,G)$, let us make some remarks.

In fact, the definition of functorial TQFT can be extended so that one deals not only with manifolds of codimension 0 and 1, but also with manifolds of arbitrary codimension, even a point. One should assign a $(k-1)$-category to a manifold of codimension $k$.

For a physical interpretation, for a $d$-dimensional extended functorial TQFT $Z$, $Z(S^{d-1})$ is the space of local operators essentially because $S^{d-1}$ is a link of a point in $\R^d$. By the same logic, $Z(S^{d-2})$ is the 1-category of line operators, $Z(S^{d-3})$ is the 2-category of surface operators, and so on. Similarly, $Z(\operatorname{pt})$ is the $(d-1)$-category of boundary conditions because a point is a link of the boundary $\R^{d-1}$ of a half-plane $\mathbb{H}^d$. In fact, the cobordism hypothesis says that $Z(\operatorname{pt})$, if it exists, determines everything about the given TQFT $Z$.

In the context of mirror symmetry, where one considers a $d=2$ TQFT, one may imagine there exist two $d=2$ extended TQFTs, say $Z^{d=2}_{X,A}$ and $Z^{d=2}_{X^\vee,B}$, and an equivalence of categories between the categories of boundary conditions $Z^{d=2}_{X,A}(\operatorname{pt})$ and $Z^{d=2}_{X^\vee,B}(\operatorname{pt})$ determines everything there is. Indeed, Kontsevich's homological mirror symmetry proposal, which is an equivalence between a Fukaya category and a derived category of coherent sheaves, is precisely of this format.\footnote{Enumerative mirror symmetry, which is alluded above, is also believed to follow from this categorical statement, upon making some additional choice as dictated by physics.}

It is then natural to ask if one can find such a higher structure in the context of $d=3$, $\mathcal{N}=4$ theories as well. However, there is a salient point to be addressed. The objects we deal with are the ones of interest in geometric representation theory. Most of the constructions in geometric representation theory are algebraic, and from the perspective of physics, this algebraicity is to be imposed even on the spacetime manifold $M$. This may seem incompatible with either physicists' TQFT or functorial TQFT; what exactly do we mean by $S^2$ or $S^1$ in an algebraic world? \phil{this is precisely what my work with Chris did but maybe it is too much.}

Let us summarize what is known in the two worlds for both the 3d A-model and 3d B-model:
\begin{center}
\begin{tabular}{|c|c|c|}
\hline
& A-model & B-model\\
\hline
$Z(S^2)$ & ? &  \multirow{2}{*}{known} \\	
$Z(S^2_\operatorname{dR})$ & [BFN] & \\	
\hline
$Z(S^1)$ & ? &  ? \\	
$Z(S^1_{\operatorname{dR}})$ & [HY] & [HY] \\	
\hline
$Z(\operatorname{pt})$ & ? & ? \\	
\hline
\end{tabular}
\end{center}
\begin{itemize}
	\item In the 3d B-model, the Betti and de Rham model for the space $Z_{\operatorname{B}}(S^2)$ of local operators should coincide, essentially because $S^2$ is contractible.
	\item In the 3d A-model, the BFN construction of the Coulomb branch is based on the de Rham model, and we still miss the Betti construction of the Coulomb branch.
	\item The de Rham model for the category of line operators for both the A-model and the B-model is suggested by [HY] in conjunction with relations with the geometric Langlands program.
	\item In simple cases, Betti constructions are suggested by \cite{gammagePerverseSchobers2022}.
\end{itemize}}

\section{Higgs and Coulomb branches}
\label{sec:Higgs-and-Coulomb}

This section focuses on the Coulomb and Higgs branches in one particularly important case:  the $d=3$, $\mathcal{N}=4$ SUSY $\sigma$-model into $\mathbb{H}^n$\notices{,}\arxiv{ (often called hypermultiplets),}gauged by the action of a subgroup $G_c\subset U(n,\mathbb{H})$.  The fields corresponding to the map to $\mathbb{H}^n$ are often called the ``matter content'' of the theory.
It is often more convenient to forget the coordinates on $\mathbb{H}^n$ and think of it as a general $\mathbb{H}$-module $X$ with a choice of norm and an action of $G_c$.  It will also simplify things for us to consider $X$ as a $\C$-vector space with complex structure $I$ and the induced action of the complexification $G$ of $G_c$; we can encode the action of the quaternions $J$ and $K$ in the holomorphic symplectic form $\Omega(x,y)=\langle Jx,y\rangle +i\langle Kx,y\rangle$. 
We will denote the corresponding theory by $\mathcal{T}(X,G)$ and denote the Higgs and Coulomb branches by $\mathfrak{M}_{\operatorname{A/B}}(X,G)$.  

Both of these varieties have concrete mathematical descriptions, which we will describe here as best we can in limited space.  Both can be derived from manipulations in infinite-dimensional geometry, using the principle that the Hilbert space of a physical theory is obtained by geometric quantization of the phase space of the theory.  This geometric quantization is easiest if $X=T^*N$ for a $G$-representation $N$. In incredibly rough terms, this phase space comes from maps of $S^2$ into the cotangent bundle of the quotient $N/G$ satisfying certain properties. These are easiest to explain if we deform our $S^2$ to be the boundary of the cylinder  \[\mathsf{Cyl}_\delta=\{(x,y,z) \mid x^2+y^2\leq \delta, |z|\leq \delta\}\] 
\begin{wrapfigure}{R}{.14\textwidth}
\begin{tikzpicture}[scale=.5,MyPersp,font=\large]
	\def\h{1.5}
    \def\a{-50}
\draw[->] (0,0,{.5*\h}) -- ({2*cos(\a)},{2*sin(\a)},{.5*\h}) node[left] {$x$};
\draw[->] (0,0,{.5*\h}) -- ({2*sin(\a)},{-2*cos(\a)},{.5*\h}) node[right] {$y$};
\draw[->] (0,0,{.5*\h}) -- (0,0,{1.3*\h}) node[above] {$z$};
		\fill[blue,fill opacity=.05]  
		 (1,0,{\h})--(1,0,0)
		\foreach \t in {0,-2,-4,...,-180}
			{--({cos(\t)},{sin(\t)},0)}
-- (-1,0,0)--(-1,0,{\h})
		\foreach \t in {180,178,...,0}
			{--({cos(\t)},{sin(\t)},{\h})}--cycle;
		\draw[gray,   very thick] (1,0,0)
		\foreach \t in {0,2,4,...,180}
			{--({cos(\t)},{sin(\t)},0)};
		\fill[blue!20!white,fill opacity=.5]
		 (1,0,{\h})--(1,0,0)
		\foreach \t in {0,-2,-4,...,-180}
			{--({cos(\t)},{sin(\t)},0)}
-- (-1,0,0)--(-1,0,{\h})
		\foreach \t in {-180,-178,...,0}
			{--({cos(\t)},{sin(\t)},{\h})}--cycle;
\draw[gray] (1,0,0)--(1,0,{\h});
\draw[gray] (-1,0,0)--(-1,0,{\h});
	\draw[gray, very thick] (1,0,0) 
		\foreach \t in {0,-2,-4,...,-180}
			{--({cos(\t)},{sin(\t)},0)};
	\draw[gray, very thick] (1,0,\h) 
		\foreach \t in {2,4,...,360} 
			{--({cos(\t)},{sin(\t)},{\h})}--cycle;
                      \end{tikzpicture}
\end{wrapfigure}
for some real number $\delta>0$. 

We will frequently refer to the top, bottom, and sides of this cylinder, by which we mean the unit disks in the $z=\delta,-\delta$ planes, and the portion of the boundary in between.
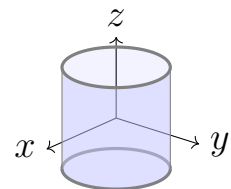
\begin{enumerate}\renewcommand{\theenumi}{\Alph{enumi}}
     \item \label{fns:A} The algebra $\alg_{\operatorname{A}}$ is the algebra of {\it locally constant} functions on the space of maps of the cylinder to $N/G$ which are constant on the sides and {\it holomorphic} on the top and bottom.
    \item \label{fns:B} The algebra $\alg_{\operatorname{B}}$ is the algebra of {\it holomorphic} functions on the space of maps of the cylinder to $N/G$ which are constant on the sides and {\it locally constant} on the top and bottom.
\end{enumerate}
We have phrased this to emphasize the parallelism, that is, how the difference between  the A- and B-twists is reflected by the placement of  ``locally constant'' and ``holomorphic.''  In the sections below, we will unpack more carefully how we interpret the concepts in the formulations (\ref{fns:A}) and (\ref{fns:B}), since some generalization is necessary.

\subsection{Higgs branches}

First, we consider the B-twist.  While second in alphabetical order, the associated Higgs branch is easier to precisely understand, and thus generally attracted more attention in the mathematical literature.  According to the description (\ref{fns:B}), the algebra $\alg_{\operatorname{B}}$ should be functions on constant maps $S^2\to N/G$.  Here 
in addition to a point in $N/G$, one should also consider a covector to this quotient (see \cite[\S 7.14]{bravermanCoulombBranches2018}).

We can define this more concretely using the moment map 
$\mu\colon X\to \mathfrak{g}^*$ of $G$ on the symplectic vector space $(X,\Omega)$. If $X=T^*N$, then $\mu^{-1}(0)$ consists of all pairs of $n\in N$ and covectors $\xi$ that vanish on the tangent space to the orbit through $n$ (and thus can be considered covectors on the quotient).   
\begin{definition}
	The Higgs branch $\mathfrak{M}_{\operatorname{B}}(X,G)$ is defined as a {\bf holomorphic symplectic quotient}, that is, one has $\alg_{\operatorname{B}}(X,G)=Z_{\operatorname{B}}(S^2)=\C[\mu^{-1}(0)]^G$, the complex polynomial functions on $\mu^{-1}(0)$ which are $G$-invariant, and $\mathfrak{M}_{\operatorname{B}}=\Spec\alg_{\operatorname{B}}$.  The points of this space are in bijection with closed $G$-orbits in $\mu^{-1}(0).$
\end{definition}

The resulting variety is typically singular symplectic, since we are applying a version of symplectic reduction.
	One can reasonably ask if this variety has a symplectic resolution. This does not happen in all cases, perhaps not even in most cases, but in some cases it does.
	
There are two particular examples that we will focus on in this article: abelian and quiver gauge theories. In both cases, the target space $X=T^*N$ is the cotangent bundle of a  $\C$-representation $N$ of $G$.

\subsubsection{Abelian/hypertoric gauge theories}

Assume that $G$ is abelian.  Since it is connected and reductive (i.e. the complexification of a compact group), this means $G\cong (\C^{\times})^k$ for some $k$.  For any $\C$-representation $N$ of $G$, we can choose an isomorphism $N\cong \C^n$ such that $G\hookrightarrow D$ is a subgroup of the full group $D$ of $n\times n$ diagonal matrices (that is, we choose a weight basis). 

These ingredients are typically used in the construction of a toric variety: the symplectic reduction of the symplectic vector space $(N,\omega_I)$ or equivalently, GIT quotient $N/\!\!/_{\lambda}G$, at any regular value of the moment map will give a quasi-projective toric variety for the action of the quotient $F=D/G$.  

The construction of the Higgs branch $\mathfrak{M}_{\operatorname{B}}(T^*N,G)$ of this theory is thus a quaternionic version of the construction of toric varieties.  The resulting variety is called a {\bf hypertoric variety} or {\bf toric hyperk\"ahler variety}.  This variety has complex dimension $2(\dim N-\dim G)$.  Probably the most familiar examples for readers are the following: 
\begin{ex}\label{ex:TPn}
    Let $G=\{\varphi I \mid \varphi \in \C^{\times}\}\subset D$ be the scalar $n\times n$ matrices.  We can consider the elements of $T^*N$ as pairs of an $n\times 1$ column vector  $\mathbf{a}$ and a $1\times n$ row vector $\mathbf{b}$, with the group $G\cong \C^{\times}$ acting by 
\[\varphi \cdot (\mathbf a,\mathbf b)=(\varphi\mathbf a,\varphi^{-1} \mathbf b) .\]   The outer product $\mathbf{ab}$ is thus an $n\times n$ matrix of rank $\leq 1$, invariant under the action of $G$. Thus, $(\mathbf a,\mathbf b)\mapsto \mathbf{ab}$ defines a map $\mathfrak{M}_{\operatorname{B}}(T^*\C^n,G)\to M_{n\times n}(\C)$.   The moment map $\mu(\mathbf a,\mathbf b)=\mathbf{ba}$ is defined by the dot product, so $\mu(\mathbf a,\mathbf b)=0$ if and only if $\mathbf{ab}$ is nilpotent.  Thus, we have a map $\mathfrak{M}_{\operatorname{B}}(T^*\C^n,G)\to \rkone$ to the space of nilpotent matrices of rank $\leq 1$. 

The  $G$-orbit through $(\mathbf a,\mathbf b)$ is closed if and only if $\mathbf a$ and $\mathbf b$ are both non-zero or both zero; you can see from this that the map above is a bijection. Thus we find $\mathfrak{M}_{\operatorname{B}}( T^*\C^n,G) \cong  \rkone$.
\end{ex}
\begin{ex}\label{ex:CZn}
Let  $G\subset D$ be the diagonal matrices of determinant 1.  We can again think of $T^*N$ as pairs $(\mathbf a,\mathbf b)$.  In this case, the moment map condition guarantees that $a_ib_i=a_jb_j$ for all $i,j$, and the closed orbit condition that if $a_i=0$ for some $i$, then $a_j=0$ for all other $j$, and similarly with $b_*$'s. 
By multiplying with a diagonal matrix, we can assume that $a_1=\cdots =a_n$ and $b_1=\cdots =b_n$ as well.  This defines a surjective map $\C^2 \to \mathfrak{M}_{\operatorname{B}}$, sending $(x,y)$ to $\mathbf{a}=(x,\dots, x)$ and $\mathbf{b}=(y,\dots,y)$.  However, this is not injective: the diagonal matrices $e^{2\pi ik/n} I$ have determinant 1 and define an action of the cyclic group $\Z_n$ on the image of this map from $\C^2$.  This matches the action of the matrix  $\operatorname{diag}
	(e^{2\pi ik/n},e^{-2\pi ik/n})$ on $\C^2.$  Thus, we find $\mathfrak{M}_{\operatorname{B}}(T^*\C^n,G)\cong \C^2/\mathbb{Z}_{n}$.  
\end{ex}

While the hypertoric varieties for other tori are less familiar and more complicated, they still have a very combinatorial flavor, and typically questions about them can be reduced to studying hyperplane arrangements, much as toric varieties can be studied using polytopes.  Notably, they all possess symplectic resolutions, constructed with GIT quotients or equivalently hyperhamiltonian reduction at non-zero moment map values.  For Example \ref{ex:TPn} above, this resolution is $T^*\mathbb{CP}^{n-1}$ and for Example \ref{ex:CZn}, it is the unique crepant resolution obtained by iterated blowups at singular points. 

\subsubsection{Quiver gauge theories}

The most famous examples of these reductions are {\bf Nakajima quiver varieties}.  These appear when $N$ is the space of representations of a quiver on a fixed vector space.  That is, we fix a directed graph $\Gamma$, and a pair of vectors $\Bv,\Bw$ whose components are indexed by the vertex set $I$.  The group $G=\prod_{i\in I} \operatorname{GL}(v_i;\C)$ has representations $\C^{v_i}$ for each $i\in I$.  The representation we will consider is 
\[N= \Big(\bigoplus_{i\to j}\Hom(\C^{v_i},\C^{v_j})\Big)\oplus \Big(\bigoplus_{i\in I} \Hom(\C^{v_i},\C^{w_i})\Big).\] 
We want a left group action, so $(A,B)\in \operatorname{GL}(v_i;\C)\times \operatorname{GL}(v_j;\C)$ acts on $M\in \Hom(\C^{v_i},\C^{v_j})$ by $BMA^{-1}$.  We call gauge theories $\mathcal{T}(T^*N,G)$ for this choice of $G$ and $N$ {\bf quiver gauge theories}, and the Higgs branches $\mathfrak{M}_{\operatorname{B}}(T^*N,G)$ are Nakajima quiver varieties.  These are geometric avatars of the $\mu$ weight space of a representation of highest weight $\lambda$ for the Kac--Moody algebra with Dynkin diagram $\Gamma$.  These are characterized by  $\alpha_i^{\vee}(\lambda)=w_i$ (that is, $w_i$ is its weight for the $\mathfrak{sl}_2$ corresponding to the node $i$), and $\lambda-\mu=\sum v_i\alpha_i$ (that is, the $v_i$'s tell us how far down from the highest weight space we have moved).  

We can think of the points of $N$ as ``framed quiver representations,'' that is, as an assignment of:
\begin{enumerate}
    \item a vector space $\C^{v_i}$ to each node of $i$, and
    \item a framing map $\C^{v_i}\to \C^{w_i}$ for each node, and
    \item  a linear map $\C^{v_i} \to \C^{v_j}$ to each oriented edge $i\to j$.
\end{enumerate}  
The action of $G$ is by changing basis in $\C^{v_i}$, so $G$-orbits in $N$ correspond to isomorphism classes of framed quiver representations\footnote{The definition of the category of framed quiver representations that makes this statement correct is left as an exercise to the reader.}.
Physicists will typically draw two copies of each node, one in a circle filled with $v_i$, and one in a square filled with $w_i$, and draw in the edges of $\Gamma$ between the first copies, and then edges between the circle and square copies of vertex (not drawing vertices with 0's).
We can then interpret $T^*N$ as representations of the doubled quiver.

\begin{ex}\label{ex:Gr}
  If we have a quiver with a single vertex so that we have a single $v$ and $w$, then $N=\Hom(\C^v,\C^w)$, and $X=T^*N=\Hom(\C^v,\C^w)\oplus \Hom(\C^w,\C^v)$, that is a pair of matrices $(A,B)$ which are $w\times v$ and $v\times w$ respectively.  The moment map in this case is 
$\mu(A,B)=BA$. If $v=1$, then this reduces to Example \ref{ex:TPn}; more generally, the matrix $AB$ is unchanged by the action of the $\operatorname{GL}(v)$, and defines an isomorphism 
\[\mathfrak{M}_{\operatorname{B}}(T^*N,G)\cong \{C\in M_{w\times w}(\C) \mid C^2=0,\ \operatorname{rk}(C)\leq v\}.\]
\end{ex} 
 
\maybelater{This shows that if $\lambda\neq 0$, then $\mathfrak{M}_{\!\alpha, \beta,\gamma}$ is isomorphic to the coadjoint orbit of matrices with eigenvalue $\lambda$  with  multiplicity 1 and 0 with (geometric) multiplicity $n-1$, via the map above.  

On the other hand, assume $\lambda=0$ and that $\alpha >0$.  In each $\operatorname{GL}(1;\C)$-orbit where $b\neq 0$, there is a unique $U(1)$-orbit of vectors that satisfy $|b|^2 -|a|^2=\alpha $; of course if $b=0$, then there are no such solutions.  Thus, we have a map $\mathfrak{M}_{\!\alpha, 0,0}\to \mathbb{CP}^{n-1}$ sending the orbit $(a,b)$ to the line spanned by $b\neq 0$.  The fiber of this map is given by $\{a\in \C^{w} \mid ba=0\}$; one can check that this is the fiber of the cotangent bundle to $\mathbb{CP}^{n-1}$, so $\mathfrak{M}_{\!\alpha, 0,0}\cong T^*\mathbb{CP}^{n-1}$.    For higher values of $v$, we can get the cotangent bundles of Grassmannians by similar logic, with slightly trickier linear algebra.  }

Other important examples:
\begin{enumerate}
    \item[E1.] This quiver gives the space of $n\times n$ nilpotent matrices:
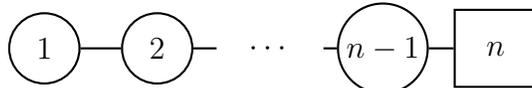
\begin{equation*}
\tikz{
	\node[draw, thick, circle, inner sep=6pt,fill=white] (a) at (-4.5,0) {$1$};
	\node[draw, thick, circle, inner sep=6pt,fill=white] (b) at (-3,0) {$2$};	
		\node[draw, thick, circle, inner sep=2pt,fill=white] (d) at (0,0) {$n-1$};
	\node[draw, thick,inner sep=12pt] (s) at (1.5,0){$n$};
	\node[inner sep=12pt,fill=white] (c) at (-1.5,0){$\cdots$};
		\draw[thick] (a) -- (b) ;
		\draw[thick] (b) -- (c) ;
		\draw[thick] (c) -- (d) ;
		\draw[thick] (d) -- (s) ;
}
\end{equation*}

\item[E2.] This quiver gives the symmetric power $\Sym^n\C^2=(\C^2)^n/S_n$.
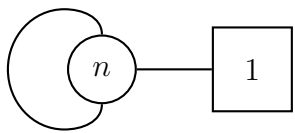
\begin{equation*}
\tikz{
		\node[draw, thick, circle, inner sep=6pt,fill=white] (d) at (0,0) {$n$};
	\node[draw, thick,inner sep=12pt] (s) at (2,0){$1$};
		\draw[thick] (d.north) to[out=90,in=0] (-.5,.8) to[out=180,in=90] (-1.25,0) to[out=-90,in=180] (-.5,-.8) to[out=0,in=-90] (d.south) ;
		\draw[thick] (d) -- (s) ;
}\end{equation*}

\end{enumerate}

There are many variations on these, but these will suffice as our main examples for the rest of this article.  \maybelater{If you are of a more algebraic bent, it might be more interesting to replace these with their corresponding quantizations, which we can construct by replacing $T^*N$ with the differential operators on $N$ and doing a non-commutative Hamiltonian reduction:
\begin{enumerate}
    \item The quantization of the nilpotent cone is the universal enveloping algebra $U(\mathfrak{sl}_n)$.
    \item The quantization of the symmetric power is the spherical Cherednik algebra of the Coxeter group $S_n$.
\end{enumerate}}
All of these examples have symplectic resolutions obtained by replacing the affine quotient with a GIT quotient: the space of rank $\leq v$ matrices is resolved by $T^*\operatorname{Gr}(w,v)$ if $v\leq w/2$, the nilpotent cone is resolved by the cotangent bundle of the flag variety (this is a special case of the Springer resolution), and the symmetric power $\Sym^n\C^2$ is resolved by the Hilbert scheme of $n$ points on $\C^2$.  \arxiv{All Nakajima quiver varieties that satisfy reasonable technical conditions have such resolutions.}

\subsection{Coulomb branches}
Compared to Higgs branches, Coulomb branches are  harder to describe.  In older papers, one will generally see the statement that the 
``classical''  Coulomb branch is $T^*{}^LT/W$ for ${}^LT$ Langlands dual to the maximal torus $T$ of $G$.  However, this is not the true answer, as there are non-trivial ``quantum corrections.'' In certain special cases, the true Coulomb branch could be determined by other methods:
\begin{itemize}
	\item Work of Hanany--Witten \cite{hananyTypeIIB1997} and  extensions identified the Coulomb branches of the quiver gauge theories E1. and E2. using string dualities.
	\item It is implicit in \cite{telemanGaugeTheory2014} that the Coulomb branch of a pure gauge theory (meaning $N=0$) with gauge group $G$ is the (Lie algebra) regular centralizer variety for $G^{\vee}$, called the Bezrukavnikov--Finkelberg--Mirkovi\'c space $\operatorname{BFM}(G^{\vee})$ there.  
 \arxiv{This space is identified with the spectrum of the equivariant homology of the affine Grassmannian of $G^{\vee}$ in \cite{bezrukavnikovEquivariantHomology2005}.}
\end{itemize}

Work of Braverman, Finkelberg, and Nakajima \cite{BFN}
gives an explicit mathematical definition of the Coulomb branch based on the geometry of affine Grassmannians when $X=T^*N$.  Let us try to roughly explain the source of this construction, which can look quite intimidating.  

By the description (\ref{fns:A}), we should consider maps to $N/G$ that are holomorphic on the top and bottom disks of the cylinder and constant along the sides. We will shrink the parameter $\delta$ that defines the cylinder to be infinitesimally small and only consider the Taylor expansion at the origin in the top and bottom planes $z=\pm \delta$. If we let $t = x + iy$, then each holomorphic map $D\to N/G$ corresponds to a Taylor series $n(t)\in N[[t]]$, and two give the same map if they are in the same orbit of the $G$-valued Taylor series $G[[t]]$.

Finally, the map should be constant along the sides of the cylinder.  Mapping to a quotient means that two things are equal if they are in the same orbit for a group-valued function on the circle.  Since this is only on the sides of the cylinder, the group-valued function comparing the top and bottom might have a pole at the origin.  Thus, if $n_\pm(t)$ is the Taylor expansion at $(0,0,\pm \delta)$, then we must have $g_{\pm} n_{-}(t)=n_+(t)$ for some $G$-valued Laurent series $g_{\pm}(t)\in G((t))$.  

\begin{definition}
    The {\bf BFN space} for $(G,N)$ is the quotient of the set $\{(n_+,n_-,g_{\pm}) \mid g_{\pm} n_{-}(t)=n_+(t)\}$ by the action of $G[[t]]\times G[[t]]$ given by \[(h_+,h_-)\cdot (n_+,n_-,g_{pm})=(h_+n_+,h_-n_-,h_+g_{\pm}h_-^{-1}) \]
\end{definition}
By (\ref{fns:A}), we should consider the locally constant functions on this space.  We have to be careful, and in fact, we need to consider the Borel--Moore homology (very carefully defined) of this quotient.  That is:
\begin{definition}\label{def:Coulomb}
    The algebra $\alg_{\operatorname{A}}(T^*N,G)$ is the Borel--Moore homology of the BFN space.
\end{definition}
While the action of $G[[t]]\times G[[t]]$ is not free, we can still interpret the homology of the quotient using equivariant topology; the interested reader should refer to \cite{BFN} for a more precise discussion.  We can still compute using usual methods from finite-dimensional topology and, in particular, identify the Coulomb branches in many cases.  

\arxiv{If $X$ is a symplectic representation of $G$ which cannot be written as $X=T^*N$ for $N$ an invariant subspace, this definition must be modified, and it seems there is an obstruction to the existence of a Coulomb branch.  This approach has recently been developed in work of  Braverman--Dhillon--Finkelberg--Raskin--Travkin \cite{bravermanCoulombBranches2022} and of Teleman \cite{telemanCoulombBranches2022}.}

The Coulomb branch comes equipped with a $\C^{\times}$ action, induced by the homological grading on Borel--Moore homology.  Unfortunately, there can be operators of negative degree.  We call such theories {\bf bad}.  Most other cases are called {\bf good}\arxiv{ (though there is a third possibility, which as the reader can probably guess is ``ugly,'' but we will ignore this case).}

One bad example which had already attracted the interest of mathematicians was the case of pure gauge theory where $N=0$.  The Higgs branch is quite degenerate in this case, but $\mathfrak{M}_{\operatorname{A}}=\operatorname{BFM}(G)$ is the phase space of the rational Toda lattice, that is, the universal (Lie algebra) centralizer, restricted to the Kostant slice \cite{bezrukavnikovEquivariantHomology2005}; see \cite[\S 5.1]{telemanGaugeTheory2014} for a longer discussion of this variety and references on it. The other interesting cases we know fall into the two cases discussed before: 

\subsubsection{Abelian/hypertoric gauge theories}

If $G$ is an abelian group and acts faithfully on $N$, then the Coulomb branch will coincide with the Higgs branch of another good abelian theory.\arxiv{  If the action isn't faithful, the theory will be ugly.}
 
Recall that we can assume that $G\subset D$ is a subgroup of the group $D$ of diagonal matrices. The Langlands dual group ${}^LF$ of $F=D/G$ can be realized as the connected subgroup of $D$ whose Lie algebra is the perpendicular to $\mathfrak{g}\subset \mathfrak{d}=\C^n$; this is also isomorphic to $(\C^{\times})^k$, but the isomorphism $F\cong {}^LF$ is not canonical.  
We thus have two Langlands dual short exact sequences of tori:
\begin{align*}
	1\to G\to & D \to F \to 1\\
	1\to {}^LG\leftarrow & D\leftarrow {}^L F \leftarrow 1.
\end{align*}
\arxiv{There is a relationship between the corresponding gauge theories:}
\begin{theorem}
  We have isomorphisms $\mathfrak{M}_{\operatorname{A/B}}(T^*\C^n,G)\cong \mathfrak{M}_{\operatorname{B/A}}(T^*\C^n,{}^LF)$. 
\end{theorem}
\arxiv{This isomorphism was widely expected before a precise definition of Coulomb branches was available.  It seems to have first been checked using the mathematical definition by Dimofte and Hilburn; see \cite[\S 3.3]{bullimoreCoulombBranch2017} for a physical discussion of this isomorphism, and \cite[\S 4(vii)]{BFN} for an elegant mathematical proof.

While a full description of this isomorphism is outside our scope here, let us give a flavor of it. First, both sides carry an action of $F$, or on $\alg_{\operatorname{A/B}}$ a grading by the weight lattice of $F$, or equivalently, the coweight lattice of ${}^LF$:
\begin{enumerate}
	\item On $\mathfrak{M}_{\operatorname{B}}(T^*\C^n,G)$, this is induced by the $D$-action on $T^*\C^n$.
	\item The desired grading on $\alg_{\operatorname{A}}(T^*\C^n,{}^LF)$ is induced by the bijection of components of the affine Grassmannian of ${}^LF$ to the coweight lattice. 
\end{enumerate}
Second, in both algebras, the degree 0 elements form a copy of $\Sym(\mathfrak{f})$: 
\begin{enumerate}
	\item On $\mathfrak{M}_{\operatorname{B}}(T^*\C^n,G)$, these are the polynomial functions that factor through the moment map $\mathfrak{M}_{\operatorname{B}}(G,T^*\C^n)\to \mathfrak{f}^*$.
	\item In $\alg_{\operatorname{A}}(T^*\C^n,{}^LF)$, the degree 0 elements are the Borel--Moore homology of a point modulo the action of ${}^LF[[t]]$, which is the same as $\Sym(\mathfrak{f})\cong H^*_{{}^LF}(\operatorname{pt})$.  
\end{enumerate}
The isomorphism of algebras is close to being determined by matching these two aspects of the Higgs and Coulomb branches.}
\notices{This isomorphism was widely expected earlier, but seems to have first been checked using the mathematical definition by Dimofte and Hilburn; see \cite[\S 3.3]{bullimoreCoulombBranch2017} for a physical discussion of this isomorphism, and \cite[\S 4(vii)]{BFN} for an elegant mathematical proof.

Let us give a sketch of it. First, the algebras $\alg_{\operatorname{A/B}}$ carry a grading by the weight lattice of $F$/coweight lattice of ${}^LF$: (1) On $\alg_{\operatorname{B}}(T^*\C^n,G)$, induced by the $D$-action on $T^*\C^n$;
	(2) On $\alg_{\operatorname{A}}(T^*\C^n,{}^LF)$, induced by the bijection of components of the affine Grassmannian of ${}^LF$ to the coweight lattice. 
Second, in both algebras, the degree 0 elements form a copy of $\Sym(\mathfrak{f})$: 
(1) On $\mathfrak{M}_{\operatorname{B}}(T^*\C^n,G)$, the polynomial functions that factor through the moment map $\mathfrak{M}_{\operatorname{B}}(G,T^*\C^n)\to \mathfrak{f}^*$;
	(2) In $\alg_{\operatorname{A}}(T^*\C^n,{}^LF)$, the Borel--Moore homology of a point modulo the action of ${}^LF[[t]]$, isomorphic to $\Sym(\mathfrak{f})\cong H^*_{{}^LF}(\operatorname{pt})$.  
The isomorphism of algebras is close to being determined by matching these two aspects of the Higgs and Coulomb branches.}

\subsubsection{Quiver gauge theories}
\label{sec:quiver-gauge-theories}

The other class of theories we discussed before, and the richest we currently understand, are the quiver gauge theories.  The resulting Coulomb branches are thus the mirrors of the Nakajima quiver varieties. Those which are good in the sense discussed above correspond to pairs of a highest weight $\lambda$ for the Kac--Moody Lie algebra with Dynkin diagram given by the quiver, and a dominant weight $\mu$.  These are characterized by:
\begin{enumerate}
    \item The highest weight vector has weight $w_i$ for the root $\mathfrak{sl}_2$ for the node $i$ (that is, $\alpha_i^{\vee}(\lambda)=w_i$).
    \item For simple roots $\alpha_i$, we have $\lambda-\mu=\sum v_i\alpha_i.$
\end{enumerate}
When $\Gamma$ is an ADE Dynkin diagram, these Coulomb branches have an interpretation in terms of the affine Grassmannian of the finite dimensional group $G_{\Gamma}$ whose Dynkin diagram is $\Gamma$. 
We can interpret $\lambda$ and $\mu$ as coweights of $G_{\Gamma}$, and thus consider the closure of the orbit  $\overline{\mathsf{Gr}}^{\lambda} =\overline{\GO t^{\lambda}\GO/\GO}\subset \GK/\GO$.
\begin{proposition}[\mbox{\cite[Th. 3.10]{BFNplus}}]
	The Coulomb branch $\mathfrak{M}_{\operatorname{A}}(T^*N,G)$ for an ADE quiver gauge theory is isomorphic to the transverse slice to a generic point of $\overline{\mathsf{Gr}}^{\mu}$ inside $\overline{\mathsf{Gr}}^{\lambda}$.  
\end{proposition} 

\arxiv{This result can be extended to non-dominant weights by introducing certain generalized slices in the affine Grassmannian \cite{BFNplus}.  Since these affine Grassmannian slices are not familiar to most readers, let us discuss a few examples: }

\notices{Since these affine Grassmannian slices are not familiar to most readers, let us discuss a few examples:  }

\begin{ex}
The quivers E1., E2. both satisfy $\mathfrak{M}_{\operatorname{A}}\cong \mathfrak{M}_{\operatorname{B}}$!  This is coincidental and usually doesn't happen.  
\end{ex}
\begin{ex}
    In the case of the quiver \tikz[baseline=-2pt]{	
		\node[draw, thick, circle, inner sep=2pt,fill=white] (d) at (0,0) {$1$};
	\node[draw, thick,inner sep=4pt] (s) at (1,0){$n$};
		\draw[thick] (d) -- (s) ;
}, where $\mathfrak{M}_{\operatorname{B}}\cong \rkone$, the Coulomb branch $\mathfrak{M}_{\operatorname{A}}$ is the affine variety $\C^2/\mathbb{Z}_{n}$, with the cyclic group $\mathbb{Z}_{n}$ acting by the matrices $\operatorname{diag}
	(e^{2\pi ik/n},e^{-2\pi ik/n})$.  \arxiv{Note that in this case, the gauge group is abelian, so we can use the hypertoric description.}
\end{ex}
\begin{ex} In the quiver below, the Higgs and Coulomb branches are reversed from the previous example: $\mathfrak{M}_{\operatorname{B}}\cong \C^2/\mathbb{Z}_{n}$, $ \mathfrak{M}_{\operatorname{A}}\cong \rkone$. \arxiv{Again, this case is hypertoric and dual to the previous one.}
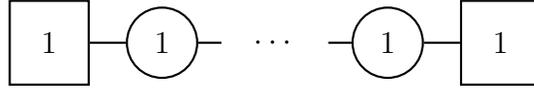
\begin{equation*}
\tikz{
	\node[draw, thick,inner sep=12pt]  (a) at (-4.5,0) {$1$};
	\node[draw, thick, circle, inner sep=6pt,fill=white] (b) at (-3,0) {$1$};	
		\node[draw, thick, circle, inner sep=6pt,fill=white] (d) at (0,0) {$1$};
	\node[draw, thick,inner sep=12pt] (s) at (1.5,0){$1$};
	\node[inner sep=12pt,fill=white] (c) at (-1.5,0){$\cdots$};
		\draw[thick] (a) -- (b) ;
		\draw[thick] (b) -- (c) ;
		\draw[thick] (c) -- (d) ;
		\draw[thick] (d) -- (s) ;
}
\end{equation*}
\end{ex}

These are special cases of a much more general result.  For good quiver theories where $\Gamma$ is an affine type A Dynkin diagram (that is, a single cycle), including the Jordan quiver (a single loop), the Coulomb branch is also a Higgs branch for a theory of an affine type A Dynkin diagram, but potentially of a different size, as proven in \cite{nakajimaCherkisBow2017}.  The combinatorics of this correspondence is a little complicated, but it matches the previously known combinatorics of rank-level duality.  This suggests that the corresponding theories are mirror to each other.

\arxiv{The case of type A (i.e. linear) quivers is a special case of the cycle (setting one $v_i=0$ ``breaks'' the cycle), and in this case, the mirror will again be a quiver gauge theory for a linear  quiver.  For example, $T^*\operatorname{Gr}(w,v)$, the cotangent bundle of the Grassmannian of $v$-planes in $\C^w$, is the resolved Higgs branch of the quiver gauge theory \tikz[baseline=-2pt]{	
		\node[draw, thick, circle, inner sep=2pt,fill=white] (d) at (0,0) {$v$};
	\node[draw, thick,inner sep=4pt] (s) at (1,0){$w$};
		\draw[thick] (d) -- (s) ;
} for a single vertex, which is good if $2v\leq w$. If $2v<w$, the mirror theory is given by 
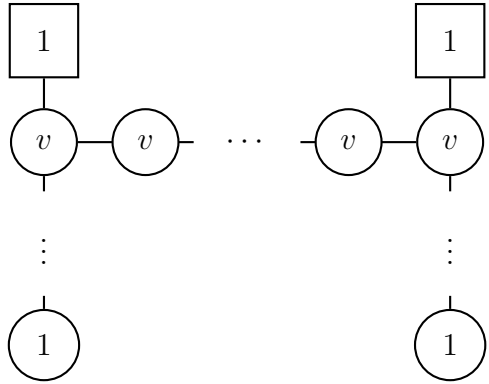
\begin{equation*}
\tikz[scale=.9]{
	\node[draw, thick, inner sep=10pt] (bb) at (-6,1.5) {$1$};
	\node[draw, thick, circle, inner sep=6pt,fill=white] (a) at (-6,-3) {$1$};
	\node[inner sep=12pt,fill=white] (ab) at (-6,-1.5){$\vdots$};
	\node[draw, thick, circle, inner sep=6pt,fill=white] (b) at (-6,0) {$v$};	
	\node[draw, thick, inner sep=10pt] (dd) at (0,1.5) {$1$};
		\node[draw, thick, circle, inner sep=6pt,fill=white] (d) at (0,0) {$v$};
	\node[draw, thick, circle, inner sep=6pt,fill=white] (e) at (0,-3) {$1$};
	\node[inner sep=12pt,fill=white] (de) at (0,-1.5){$\vdots$};
	\node[inner sep=12pt,fill=white] (c) at (-3,0){$\cdots$};
		\draw[thick] (a) -- (ab) --(b) --(bb);
		\draw[thick] (b) -- (c) ;
		\draw[thick] (c) -- (d) ;
		\draw[thick] (e) -- (de) --(d) --(dd);
				\node[draw, thick, circle, inner sep=6pt,fill=white] (d) at (-1.5,0) {$v$};
						\node[draw, thick, circle, inner sep=6pt,fill=white] (d) at (-4.5,0) {$v$};
}
\end{equation*}
with $w-2v+1$ nodes with label $v$ in the horizontal line, and if $2v=w$, then 
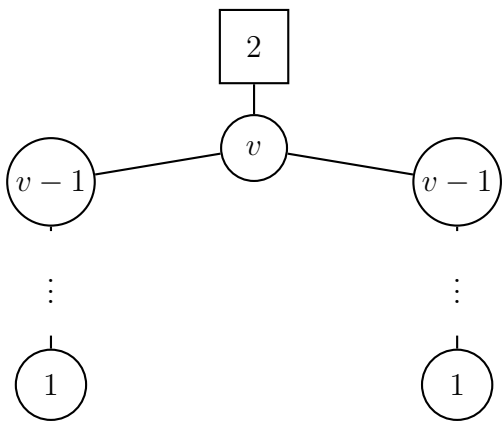
\begin{equation*}
\tikz[scale=.9]{
	\node[draw, thick, inner sep=10pt] (cc) at (-3,1.5) {$2$};
	\node[draw, thick, circle, inner sep=6pt,fill=white] (a) at (-6,-3.5) {$1$};
	\node[inner sep=12pt,fill=white] (ab) at (-6,-2){$\vdots$};
	\node[draw, thick, circle, inner sep=2pt,fill=white] (b) at (-6,-.5) {$v-1$};	
		\node[draw, thick, circle, inner sep=6pt,fill=white] (c) at (-3,0) {$v$};
		\node[draw, thick, circle, inner sep=2pt,fill=white] (d) at (0,-.5) {$v-1$};
	\node[draw, thick, circle, inner sep=6pt,fill=white] (e) at (0,-3.5) {$1$};
	\node[inner sep=12pt,fill=white] (de) at (0,-2){$\vdots$};
		\draw[thick] (a) -- (ab) --(b);
		\draw[thick] (b) -- (c) --(cc);
		\draw[thick] (c) -- (d) ;
		\draw[thick] (e) -- (de) --(d) ;
}
\end{equation*}

In this case, the combinatorics of the duality are more easily described by identifying the Higgs and Coulomb branches with the transverse slice $\mathfrak{X}^{\nu}_{\eta}$ to the orbit of nilpotent matrices with Jordan type $\eta$ in closure of the orbit with Jordan type $\nu$ for $\eta, \nu$ partitions of $n$; this is always possible.  In this case, the Coulomb branch of the same theory is the slice $\mathfrak{X}^{\eta^t}_{\nu^t}$.  In the Grassmannian case above, the Higgs branch of the theory \tikz[baseline=-2pt]{	
		\node[draw, thick, circle, inner sep=2pt,fill=white] (d) at (0,0) {$v$};
	\node[draw, thick,inner sep=4pt] (s) at (1,0){$w$};
		\draw[thick] (d) -- (s) ;
} is the closure of the nilpotent matrices of Jordan type $(2^v,1^{w-2v})$ (the image of $T^*\operatorname{Gr}(w,v)$ under the moment map to $\mathfrak{gl}_{w}^*$), whereas the Coulomb branch is the slice in the full nilpotent cone to the orbit with Jordan type $(w-v,v)$, often called a {\bf two row Slodowy slice}, which has made a number of appearances in representation theory and knot homology.  \phil{I am lost here but maybe because I didn't fully understand even the previous parts.}}

\section{Advanced directions}
\label{sec:advanced}

Having given the definition of Higgs and Coulomb branches, the reader will naturally wonder what mathematics these lead to.  There are a number of directions which are too deep to discuss in full detail, but which the interested reader might want to explore further:

\subsection{Stable envelopes}
Aganagi\'c and Okounkov \cite{aganagicEllipticStable2020}, building on earlier work of Maulik--Okounkov, define classes called {\bf elliptic stable envelopes} on each symplectic resolution with a Hamiltonian $\C^*$-action.   There are many examples of these which arise as $\mathfrak{M}_{\operatorname{A/B}}$ for different $3$d, $\mathcal{N}=4$ gauge theories.

The equivariant stable envelopes are classes in equivariant elliptic cohomology which correspond to the thimbles flowing to the different $\C^*$-fixed points on the resolution (equivalently, the stable manifolds of the real moment map, thought of as a Morse function). These play an important role in the study of enumerative geometry and are expected to be one of the key mathematical manifestations of 3d mirror symmetry.  The elliptic stable envelopes of mirror varieties are expected to be obtained from the specialization of a natural ``Mother'' class on the product $\mathfrak{M}_{\operatorname{A}}\times \mathfrak{M}_{\operatorname{B}};$ this is confirmed in the case of Example \ref{ex:Gr} by Rim\'anyi--Smirnov--Varchenko--Zhou \cite{rimanyi3dMirror2020}.  This identification switches two classes of parameters in the physical theory:
\begin{enumerate}
    \item ``masses'' which index resolutions of $\mathfrak{M}_{\operatorname{A}}$, and $\C^*$-actions on $\mathfrak{M}_{\operatorname{B}}$, and
    \item ``Fayet--Iliopoulos (FI) parameters'' which play the opposite role of indexing $\C^*$-actions on $\mathfrak{M}_{\operatorname{A}}$ and resolutions of $\mathfrak{M}_{\operatorname{B}}$. 
\end{enumerate} 

\subsection{Koszul duality of category \texorpdfstring{$\mathcal{O}$}{O}'s}

One of the mathematical phenomena which has attracted attention to 3-dimensional mirror symmetry is Koszul duality between categories $\mathcal{O}$.  These are based on a deformation quantization of the algebras $\alg_{\operatorname{A/B}}$ to non-commutative algebras.  These deformations can be understood as incorporating the action of the rotation of $\R^3$ by $S^1$ around the $z$-axis; in physics terms, this is called an $\Omega$-background.  The resulting algebra is non-commutative, since only the $z$-axis is invariant under the $S^1$-action, and two invariant points cannot switch places while staying on the $z$-axis.  
\begin{enumerate}
    \item  The algebra $\alg_{\operatorname{A}}$ is deformed by considering the $S^1$-equivariant homology of the BFN space.
    \item The algebra $\alg_{\operatorname{B}}$ is deformed by replacing $\C[X]=\operatorname{Sym} X^*$ by its Weyl algebra, which is defined by the relations $[x,y]=\hbar \Omega(x,y)$; we can replace the operations of taking the $G$-invariant functions on $\mu^{-1}(0)$ with a non-commutative analogue of Hamiltonian reduction.  
\end{enumerate}
Category $\mathcal{O}$ is a category of special modules over these non-commutative algebras.  This can be regarded as a categorification of the stable envelopes, in that instead of considering the homology classes of the thimbles flowing into fixed points, we consider sheaves of modules over a deformation quantization of $\mathfrak{M}_{\operatorname{A/B}}$ supported on these thimbles.  See \cite[\S 3]{BLPWgco} for more details.

It was noticed by Soergel that the principal block of category $\mathcal{O}$ for a semi-simple Lie algebra has an interesting self-duality property: It is equivalent to the category of ungraded modules over a graded algebra and inside the derived category $D^b(\mathcal{\tilde{O}})$ of graded modules over that algebra, there is a second ``hidden'' copy of the original category.  For variations, such as singular blocks of category $\mathcal{O}$ or parabolic category $\mathcal{O}$, a similar phenomenon occurs, but it is a copy of another category that appears; for example, the singular and parabolic properties interchange.  That is, the graded lifts of these categories are {\bf Koszul} and their Koszul dual is another category (sometimes different, sometimes the same) of a similar flavor.     

As discussed in the introduction, we can pretend that another symplectic singularity is the nilpotent cone of a new simple Lie algebra.  The definition of category $\mathcal{O}$ for a general symplectic singularity with a $\C^*$-action was given by Braden, Licata, Proudfoot, and the first author in \cite{BLPWgco}.  Computing numerous examples led these authors to the conjecture:
\begin{conjecture}
	 The categories $\mathcal{O}$ of mirror dual symplectic singularities (i.e. the Higgs and Coulomb branch of a 3d $\mathcal{N}=4$ supersymmetric gauge theory) are Koszul dual.  
\end{conjecture} A version of this conjecture (obviously, requiring more careful stating) is confirmed in \cite{websterKoszulDuality2019}.  The physical interpretation of this Koszul duality is still uncertain, though one is proposed in \cite[\S 7.5]{bullimoreBoundariesMirror2016}.

\subsection{Line operators}

Just as local operators stand for observations one can make at a single point, there are line operators that describe observations one can make along a single line. Studying these is a natural way to extend our study of 3d mirror symmetry beyond the definition of the Higgs and Coulomb branches. 

We can describe this category using the framework of $d$-dimensional extended TQFT, which assigns not just a Hilbert space to a $(d-1)$-manifold, but more generally a $k$-category to each manifold of codimension $k+1$.  We can then generalize the description of the local operators as the space $Z(S^{d-1})$ by identifying the operators supported on a $k$-plane with the $k$-category $Z(S^{d-k-1})$. 
In particular, the category of line operators should be given by $Z(S^{d-2})$.

For the 3d $\mathcal{N}=4$ theories of interest to us, these can be understood after passing to the A- or B-twisted theory. Indeed, algebraic descriptions of these categories have been proposed by Hilburn and the second author. Here we only provide a rough description of $Z(S^1)$ with a similar flavor to (\ref{fns:A}) and (\ref{fns:B}) (see \cite{bravermanCoulombBranches2018} and \cite{DGGH} for more discussion of this proposal):
\begin{itemize}
	\item In the A-twist, we obtain {\it locally constant} sheaves (that is, D-modules) on {\it holomorphic} loops $N((t))/G((t))$ in the quotient $N/G$.  
	\item In the B-twist, we obtain {\it holomorphic} sheaves (that is, quasi-coherent sheaves) on a version of the {\it locally constant} loops (that is, the small loop space) in $N/G$.  
\end{itemize}
This proposal is actually a good way to derive Definition \ref{def:Coulomb}: the trivial line is given by the pushforward D-module from $N[[t]]/G[[t]]$, and naively computing the endomorphisms of this pushforward as the Borel--Moore homology of the fiber product gives precisely Definition \ref{def:Coulomb}. 

It is an intriguing but challenging problem to identify these categories in the already known dual pairs, which one may call the \emph{de Rham 3d homological mirror symmetry} (see below for more context for the name). Recent work of Hilburn--Raskin \cite{hilburnTateThesis2022} confirms this in the case where $N=\C$, $G=\{1\}$.

\subsection{Connections to 4d field theory and the Langlands program}

Seminal work of Kapustin and Witten \cite{kapustinElectricmagneticDuality2007} interprets a version of the geometric Langlands correspondence in terms of a physical duality between 4-dimensional field theories with $\mathcal{N}=4$ supersymmetry: the supersymmetric Yang--Mills theory in 4 dimensions for a pair of Langlands dual groups are related by S-duality. Elliott and the second author \cite{elliottGeometricLanglands2018} developed a mathematical framework to describe a variant of their proposal which yields the geometric Langlands correspondence upon (categorified) geometric quantization. Moreover, by applying this procedure to the A- and B-twists of a 3d $\mathcal{N}=4$ theory, one can obtain the aforementioned categories of line operators.

This connects to the 3-dimensional perspective discussed earlier in this paper, since 3d $\mathcal{N}=4$ theories appear as boundary conditions on 4d $\mathcal{N}=4$ super Yang--Mills theory. In particular, $S$-duality of these boundary conditions, as studied by Gaiotto and Witten \cite{gaiottoSDuality2009}, is one of our most powerful tools for finding mirror theories. The theories associated to Nakajima quiver varieties for linear or cyclic (finite or affine type A) quivers arise this way, and this is the quickest route to understanding the duality of these theories discussed in Section \ref{sec:quiver-gauge-theories}.  
 
The mathematical understanding of this perspective is still an emerging topic. Hilburn and the second author proposed a new relationship between the global/local geometric Langlands program and the statement of de Rham 3d homological mirror symmetry. In independent work, Ben-Zvi, Sakellaridis, and Venkatesh realized the physical perspective in the context of the relative Langlands program and have announced, among other things, a number of interesting conjectures relating periods and special values of $L$-functions \cite{ben-zviRelativeLanglands}.

\subsection{Betti 3d mirror symmetry}

The Betti (singular) cohomology and de Rham cohomology of an algebraic variety are, of course, isomorphic, but they have different nonabelian generalizations. They manifest as the moduli spaces of local systems and of flat connections on a given variety, which are analytically isomorphic (via the Riemann--Hilbert correspondence) but algebraically different.

Most importantly for us, the complex structure on the de Rham moduli space of an algebraic curve depends on the complex structure of the underlying curve, whereas the Betti space does not. Analogously, Ben-Zvi and Nadler propose a ``Betti'' version of the geometric Langlands correspondence which gives an automorphic description of the quasi-coherent sheaves on the Betti moduli space, to complement the ``de Rham'' version of the geometric Langlands correspondence.
 
\arxiv{Analogously, Ben-Zvi and Nadler propose a ``Betti'' version of the geometric Langlands correspondence which gives an automorphic description of the quasi-coherent sheaves on the Betti moduli space \cite{ben-zviBettiGeometric2018}, to complement the ``de Rham'' version of the geometric Langlands correspondence.}

Our discussions in the earlier sections of this paper also belong to the de Rham world. Namely, in our description of moduli spaces of vacua, objects attached to $S^2$ depend on a complex structure on this curve; the appearance of structures which are holomorphic in one plane and constant on an orthogonal line is a sort of degenerate complex structure on $S^2$. On the other hand, for physicists, this perspective looks somewhat artificial, compared to treating all directions in $\R^3$ equally.

Indeed, the proposal of Kapustin and Witten \cite{kapustinElectricmagneticDuality2007} is already phrased from a Betti perspective: It does not depend on the complex structure of the curve and needs to be modified to fit with the usual (de Rham) Langlands conjecture (as is done in \cite{elliottGeometricLanglands2018}). 
Another key feature of Higgs and Coulomb branches, which is physically expected, but hard to see from a de Rham perspective, is the existence of a hyperk\"ahler metric.  These have been constructed for Higgs branches $\mathfrak{M}_{\operatorname{B}}(T^*N,G)$ using hyperhamiltonian quotients, but it is hard to imagine the construction of such a metric on the Coulomb branch in the framework of \cite{BFN}.
Possibly the most intriguing aspect of the Betti perspective is that, as it does not depend on a complex structure, it is better suited to the approach of extended TQFT. Hence, this is the framework in which one can push the approach of homological mirror symmetry to the fullest.

\notices{In the case of 2-dimensional mirror symmetry, Kontsevich made the striking realization that we can capture the equivalence of the A-model of one theory and the B-model of another as an equivalence of two triangulated (dg/$A_{\infty}$) categories: from the Fukaya category of a symplectic manifold to the derived category of coherent sheaves on a complex variety. These are the categories of boundary conditions of the respective twisted theories, and hence the equivalence of theories can be reconstructed from the equivalence of categories. In terms of extended 2d TQFT, this is an equivalence of $Z(\operatorname{pt})$'s of the dual theories.}
\arxiv{To orient the reader, recall that in the case of 2-dimensional mirror symmetry, Kontsevich made the striking realization that we can capture the equivalence of the A-model of one theory and the B-model of another expected in mirror symmetry as an equivalence of two triangulated (dg/$A_{\infty}$) categories: from the Fukaya category of a symplectic manifold to the derived category of coherent sheaves on a complex variety. These are the categories of boundary conditions of the respective twisted theories and hence the equivalence of theories can be reconstructed from the equivalence of categories.\footnote{In fact, Kontsevich also made a conjecture that the aforementioned enumerative mirror symmetry for rational curves can be deduced from this equivalence of  categories. This conjecture was further developed by Costello and is currently an active research topic being pursued by Caldararu, Tu, and their collaborators.} In terms of extended 2d TQFT, this is an equivalence of $Z(\operatorname{pt})$'s of the dual theories.}


 This provides an enticing model to follow in the 3d case. Ideally, we would assign a 2-category $Z(\operatorname{pt})$ of boundary conditions to the A- and B-twist of each theory $\mathcal{T}$ and conjecture the equivalence between those for dual theories, which we would call the \emph{Betti 3d homological mirror symmetry}. This program was put forward by Teleman \cite{telemanGaugeTheory2014}, based on a proposal of Kapustin--Rozansky--Saulina \cite{kapustinThreedimensionalTopological2009} for $\mathcal{T}(T^*N,G)$. Significant progress on the A-model 2-category has been made in the abelian case in recent work of Gammage--Hilburn--Mazel-Gee \cite{gammagePerverseSchobers2022} and Doan--Rezchikov \cite{DoanRezchikov2022} suggesting an ambitious program for a more general case.

\maybelater{
\section{Future directions}

Exactly how these dual symplectic singularities are related is still an area of active investigation, and it is beyond the scope of this article to describe in detail. \phil{I actually think moving these to the later part would be much better, again because this wouldn't make sense to readers at this point. Probably it is good to still briefly just mention at this point why 3d mirror symmetry is interesting for many other mathematicians though; like for representation theory in general (your work), for 2d mirror symmetry (Teleman?), and for geometric Langlands (Hilburn--Raskin)} Three important avenues toward this connection have appeared in the literature, though:
\begin{enumerate}
	\item 
	\item Work of the first author, Braden, Licata and Proudfoot \cite{BLPWgco} suggests that this connection manifests in a Koszul duality between categories $\mathcal{O}$, which are categories of representations supported on the same thimbles that give the elliptic stable envelopes.
	\item There are higher categories which include more and deeper information about the QFT than the Higgs and Coulomb branches.  The most approachable of these are the monoidal category of line operators in the A- and B-twists;  mathematical models of these for gauge theories have been proposed by Hilburn and Yoo (see \cite[pg. 4]{dimofteMirrorSymmetry2020} or \cite[\S 7]{bravermanCoulombBranches2018}). \phil{I think Braverman—Finkelberg’s ``related structure’’ is the right reference for this, because it came earlier and it does explain a part of what Justin and I did.} \arxiv{The hoped-for equivalence of categories of line operators is confirmed by Hilburn and Raskin in the case of the pure Yang—Mills theory of $\C^*$, whose mirror is the Yang—Mills theory with a weight 1 action on $\C$ as matter.}  The logical endpoint of this line of research would be a construction of the 2-category of boundary conditions in the A- and B-twists, \phil{this is a statement from the Betti-centric perspective, and I think one can afford distinguishing Betti vs de Rham; for instance, even if we have 2-category, that won’t subsume works on line operators; in fact, we don’t even have Betti definition of Coulomb branch.} which would play the same role as the Fukaya category and category of coherent sheaves in 2-dimensional mirror symmetry (see \cite{gammagePerverseSchobers2022} for a more detailed discussion).
\end{enumerate}  }

\printbibliography

\end{document}